\documentclass[10pt]{wlscirep}
\usepackage{graphicx}
\usepackage{graphics}
\usepackage{bm}
\usepackage{colordvi}
\usepackage{units}
\usepackage{bbm}
\usepackage{booktabs}
\usepackage{array}
\usepackage{placeins}
\usepackage{epsfig}
\usepackage{lscape}
\usepackage{changes}
\usepackage{braket}
\usepackage{leftidx}
\usepackage{pdfpages}
\title{Test of quantum thermalization in the two-dimensional transverse-field Ising model}

\author[1,*]{Benjamin Bla{\ss}}
\author[1,$\dagger$]{Heiko Rieger}
\affil[1]{Theoretical Physics, Saarland University, 66123 Saarbr{\"u}cken, Germany}

\affil[*]{bebla@lusi.uni-sb.de}
\affil[$\dagger$]{h.rieger@mx.uni-saarland.de}

\definecolor{mymagenta}{rgb}{1.0,0.0,1.0}
\definecolor{mycyan}{rgb}{0.0,1.0,1.0}
\definecolor{myyellow}{rgb}{1.0,1.0,0.0}
\definecolor{myorange}{rgb}{1.0,0.27,0.0}

\begin{abstract}
We study the quantum relaxation of the two-dimensional transverse-field Ising model after global quenches with a real-time variational Monte Carlo method and address the question whether this non-integrable, two-dimensional system thermalizes or not. We consider both interaction quenches in the paramagnetic phase and field quenches in the ferromagnetic phase and compare the time-averaged probability distributions of non-conserved quantities like magnetization and correlation functions to the thermal distributions according to the canonical Gibbs ensemble obtained with quantum Monte Carlo simulations at temperatures defined by the excess energy in the system. We find that the occurrence of thermalization crucially depends on the quench parameters: While after the interaction quenches in the paramagnetic phase thermalization can be observed, our results for the field quenches in the ferromagnetic phase show clear deviations from the thermal system. These deviations increase with the quench strength and become especially clear comparing the shape of the thermal and the time-averaged distributions, the latter ones indicating that the system does not completely lose the memory of its initial state even for strong quenches. We discuss our results with respect to a recently formulated theorem on generalized thermalization in quantum systems.
\end{abstract}

\begin{document}
\flushbottom
\maketitle

\noindent In recent years the non-equilibrium dynamics of isolated many-body quantum systems, defined by the unitary time evolution starting from a generic non-eigenstate of the Hamiltonian, has gained tremendous interest \cite{Cazalilla2006,Rigol2007,Rigol2008,Rigol2009,Manmana2007,Kollath2007,Moeckel2008,Cramer2008,Barmettler2009,Rossini2009,Schiro2010,Igloi2000,Igloi2011,Rieger2011,Blass2012,Calabrese2007,Fagotti2008,Sotiriadis2009,Calabrese2011,Calabrese2012,Calabrese2012a,Caux2013,Khatami2013,Bucciantini2014,Fagotti2014,Heyl2015,James2015,Strand2015,Essler2016}. A central point of fundamental importance is the nature of the stationary state of the observables of the system. Here one is especially interested in answering the question whether the system thermalizes, i.e. whether time-averaged observables and their probability distributions can be described by the canonical Gibbs ensemble (CGE) \cite{Kollath2007,Reimann2008,Rossini2009,Barthel2008,Biroli2010,Eckstein2008,Eckstein2009,Tsuji2013,Marcuzzi2013,Sirker2014,Gogolin2011,Riera2012,Eisert2015,Gogolin2016}:
\begin{align}
\hat{\rho}_{\text{CGE}}=\frac{1}{Z_{\text{CGE}}}e^{-\beta\hat{H}}\quad\text{with}\quad Z_{\text{CGE}}=\text{Tr}\left[e^{-\beta\hat{H}}\right]\;.
\end{align}
The Lagrange multiplier $\beta=1/T$ is the inverse temperature and is determined under the constraint of maximizing the entropy.\\
Thermalization is closely linked to the conserved quantities of the system. While in non-integrable systems only the energy is conserved, integrable systems possess additional conserved quantities which avoid thermalization. The canonical Gibbs ensemble thus cannot be applied to integrable systems, but it has been shown that their stationary state can be well described by the generalized Gibbs ensemble (GGE)\cite{Rigol2007,Rigol2008,Rigol2009,Cramer2008,Calabrese2007,Fagotti2008,Sotiriadis2009,Calabrese2011,Calabrese2012,Calabrese2012a,Cassidy2011,Caux2013,Caux2012,Mussardo2013,Pozsgay2013,Brockmann2014}, which includes all the conserved quantities of the system, the so-called charges:
\begin{align}
\hat{\rho}_{\text{GGE}}=\frac{1}{Z_{\text{GGE}}}e^{-\tilde{\beta}\hat{H}-\sum_{n}\lambda_{n}\hat{I}_{n}}\quad\text{with}\quad Z_{\text{GGE}}=\text{Tr}\left[e^{-\tilde{\beta}\hat{H}-\sum_{n}\lambda_{n}\hat{I}_{n}}\right]\;.
\end{align}
The conserved charges are taken into account by the operators $\hat{I}_{n}$ and the Lagrange multipliers $\tilde{\beta}$ and $\lambda_{n}$ are uniquely determined by the initial conditions. However new results show that care has to be taken in the definition of the GGE \cite{Wouters2014,Pozsgay2014,Pozsgay2014a,Pozsgay2014b,Goldstein2014}. While in the past only local charges of the system were considered in the GGE, recently the existence of previously unknown quasilocal charges for different field theoretical models \cite{Essler2015} as well as for the the spin-$1/2$ Heisenberg chain \cite{Ilievski2015,Ilievski2016} has been shown. These quasilocal charges have to be included in the GGE, too, to give an adequate description of the stationary state of the systems. An universal mathematical framework for the construction of the GGE including quasilocal charges has been recently formulated along with the conditions under which generalized thermalization should occur in systems of any dimensionality\cite{Doyon2016}.\\
In non-integrable systems the energy is the only conserved quantity, but non-integrability is not synonymous to thermalization. Counterexamples are the driven Rabi model \cite{Larson2013} or a non-integrable model of hard-core bosons on connected triangular lattices \cite{Hamazaki2016}. For the latter one it has been shown that the applicability of the CGE to the stationary state depends on the symmetries in the system: In case of an extensive number of local symmetries the GGE has to be applied rather than the CGE.\\
Predictions on whether non-integrable systems thermalize or not usually rely on the eigenstate thermalization hypothesis (ETH) \cite{Deutsch1991,Srednicki1994,Rigol2007,Rigol2008,Rigol2009,Rigol2012,Fratus2015,Mondaini2016}, which computes matrix elements of observables in energy eigenstates of the system. The ETH is a sufficient but not a necessary condition for thermalization and has been applied to a wide variety of systems. In contrast to this there are only few studies on thermalization in non-integrable systems for system sizes larger than those accessible with exact diagonalization which compute the time evolution of the systems. Among them are studies of the one-dimensional Bose Hubbard model with time-dependent density matrix renormalization group theory (t-DMRG) \cite{Kollath2007} and of the antiferromagnetic anisotropic Heisenberg chain with a Chebyshev polynomial expansion \cite{Konstantinidis2015,Konstantinidis2016}. In higher dimensions there are results for a $D$-dimensional effective $O(N)$-Hamiltonian close to dynamical critical points based on a renormalization-group method \cite{Chiocchetta2015,Maraga2015,Chiocchetta2016}.\\
Here we present for the first time a systematic investigation of the relaxation dynamics of a two-dimensional, non-integrable model going beyond system sizes accessible with exact diagonalization and applicable to large areas of the parameter space. We study the transverse-field Ising model in two dimensions (2D-TFIM) after global interaction quenches in the paramagnetic phase and global field quenches in the ferromagnetic phase. In contrast to the Ising chain in one dimension, the 2D model is non-integrable and cannot be solved analytically. We use a real-time variational Monte Carlo (rt-VMC) method to compute the time evolution of observables like magnetization and correlation functions with high accuracy for long time scales and large system sizes. To answer the question whether the 2D-TFIM thermalizes or not we compare the time-averaged distributions of the observables to their thermal distributions for the system in equilibrium at temperatures determined by the excess energy after the quench. For a system that thermalizes one would expect the asymptotic time-averaged distributions and the thermal distributions to be identical. We discuss our results with respect to the theorem on generalized thermalization\cite{Doyon2016}. For the interaction quenches in the paramagnetic phase, for which the conditions of the theorem are fulfilled, we indeed observe thermalization. For the field quenches on the other hand we find a continuously increasing degree of non-thermalization with increasing quench strength.

\section*{Model and methods}
\subsection*{The model}
We study the 2D-TFIM with nearest neighbour interactions on a square lattice of size $L\times L$ with periodic boundary conditions (PBC). The system is described by the Hamiltonian
\begin{align}
\hat{H}=-\frac{J}{2}\sum_{<\mathbf{R},\mathbf{R}'>}\hat{\sigma}_{\mathbf{R}}^{x}\hat{\sigma}_{\mathbf{R}'}^{x}-\frac{h}{2}\sum_{\mathbf{R}}\hat{\sigma}_{\mathbf{R}}^{z}
\end{align}
with coupling strength $J$ and external transverse field $h$. The total number of spins in the system is $N=L^{2}$, thus the dimension of the Hilbert space $\mathcal{H}$ is $2^{N}$. As basis of $\mathcal{H}$ we choose the $\mathbf{x}$-basis, in which the operator $\hat{\sigma}_{\mathbf{R}}^{x}$ measures the orientation of the spin at site $\mathbf{R}$, while $\hat{\sigma}_{\mathbf{R}}^{z}$ inverts it.\\
The model is highly symmetric. Its Hamiltonian is invariant under the global $\mathbbm{Z}_{2}$ spin flip transformation $\hat{\sigma}_{\mathbf{R}}^{x}\rightarrow-\hat{\sigma}_{\mathbf{R}}^{x}$ and $\hat{\sigma}_{\mathbf{R}}^{z}\rightarrow\hat{\sigma}_{\mathbf{R}}^{z}$ generated by the unitary operator $\hat{\Sigma}^{z}=\Pi_{\mathbf{R}}\hat{\sigma}_{\mathbf{R}}^{z}$. Due to the square lattice with PBC the Hamiltonian also shows translation, rotation and reflection symmetries. Their generators can be constructed from the unitary transposition operators $\hat{T}_{\mathbf{R},\mathbf{R}'}=\frac{1}{2}(\hat{\mathbbm{1}}+\hat{\vec{\sigma}}_{\mathbf{R}}\cdot\hat{\vec{\sigma}}_{\mathbf{R}'})$, which interchange two sites $\hat{\sigma}_{\mathbf{R}}^{x}\leftrightarrow\hat{\sigma}_{\mathbf{R}'}^{x}$ and $\hat{\sigma}_{\mathbf{R}}^{z}\leftrightarrow\hat{\sigma}_{\mathbf{R}'}^{z}$. These symmetries can also be found in the eigenstates of the Hamiltonian and are conserved under unitary time evolution.\\
Ferromagnetic long-range order exists in the thermodynamic limit ($L\to\infty$) at low temperatures and fields due to spontaneous symmetry breaking of the global spin flip symmetry and is indicated by a non-vanishing ground state magnetization in $x$-direction $\braket{\Psi_{0}|\hat{\mu}^{x}|\Psi_{0}}\neq0$ with $\hat{\mu}^{x}=\frac{1}{N}\sum_{\mathbf{R}}\hat{\sigma}_{\mathbf{R}}^{x}$. In the ferromagnetic phase the system has no energy gap in contrast to the paramagnetic phase.
\begin{figure}
\includegraphics[scale=1]{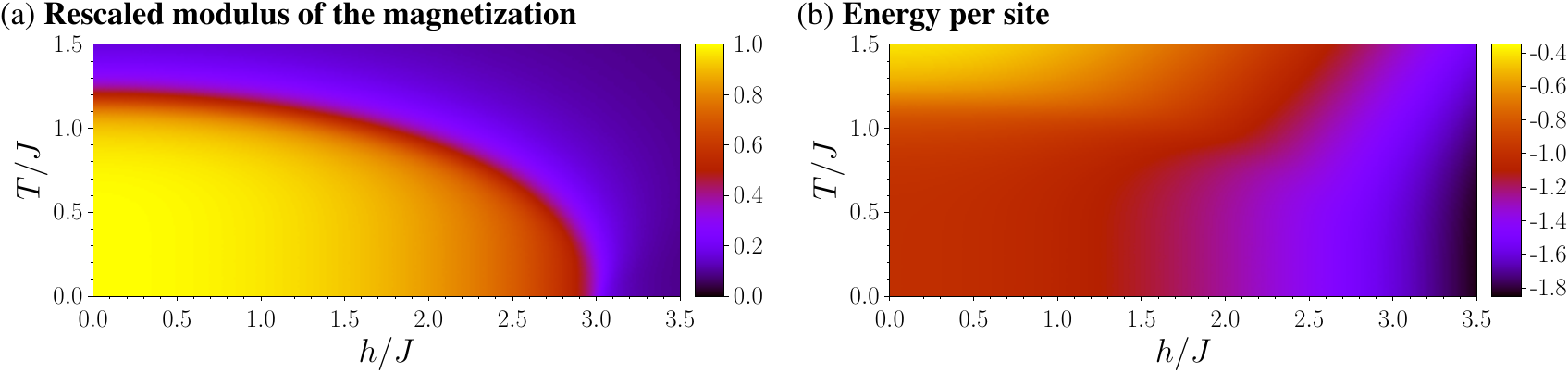}
\caption{Expectation values of (a) the rescaled modulus of the magnetization in $x$-direction $\braket{\hat{\tilde{\mu}}^{x}}_{\text{CGE}}$ according to equation (\ref{Eq:mu_tilde}) and (b) the energy per site $\frac{1}{NJ}\braket{\hat{H}}_{\text{CGE}}$ of the 2D-TFIM with PBC on a square lattice of size $16\times16$ computed with a cluster Monte Carlo algorithm in continuous imaginary time \cite{Rieger1999}. Each data point is an average over $10^{5}$ samples.}
\label{fig_phase_diagram}
\end{figure}
For $h=0$ the equilibrium phase transition occurs at $(T/J)_{\text{crit}}\approx1.135$ \cite{Onsager1944} and for $T=0$ at $(h/J)_{\text{crit}}\approx3.044$ \cite{Pfeuty1971,deJongh1998,Rieger1999}. As there is no spontaneous symmetry breaking in the finite system, here $\braket{\Psi_{0}|\hat{\mu}^{x}|\Psi_{0}}=0$ due to the global spin flip symmetry irrespective of the values of $T/J$ and $h/J$. For this reason we use for the finite systems in our numerical studies the modulus of the magnetization as order parameter. In order to get rid of finite size effects it is renormalized according to
\begin{align}
\hat{\tilde{\mu}}^{x}\equiv\frac{|\hat{\mu}^{x}|-|\mu^{x}|_{\text{min}}}{1-|\mu^{x}|_{\text{min}}}\quad\text{with}\quad|\mu^{x}|_{\text{min}}=\frac{1}{2^{N-1}}\sum_{m=0}^{N/2-1}\begin{pmatrix}N\\m\end{pmatrix}\frac{N-2m}{N}\sim\frac{1}{L}
\label{Eq:mu_tilde}
\end{align}
the expectation value of the modulus of the magnetization in the completely uncorrelated state. In the thermodynamic limit it is $\braket{\Psi_{0}|\hat{\tilde{\mu}}^{x}|\Psi_{0}}=\braket{\Psi_{0}|\hat{\mu}^{x}|\Psi_{0}}$. Figure \ref{fig_phase_diagram} (a) shows the phase diagram of $\braket{\hat{\tilde{\mu}}^{x}}_{\text{CGE}}$ for a $16\times16$ system. Already for this system size one observes only small deviations from the results for the system in the thermodynamic limit, whose critical values for $T/J$ and $h/J$ can be obtained from the Binder cumulant applying finite size scaling \cite{Binder1981}.\\
The Hamiltonian of the 1D-TFIM can be diagonalized by a transformation to a system of free fermions \cite{Pfeuty1970}. For the 2D-TFIM this is not possible as its Hamiltonian is non-local after the 2D-Jordan-Wigner transformation \cite{Fradkin1989,Wang1991,Azzouz1993}, so that there is no canonic transformation to a system of free fermions. For this reason its relaxation process after a quench cannot be described with the semiclassical theory of non-interacting quasiparticles introduced for the 1D-TFIM either \cite{Rieger2011}.

\subsection*{Quench protocol and effective temperature}
We drive the system out of equilibrium by a global quench, i.e. the system is prepared in its ground state $\ket{\Psi_{\text{i},0}}$ for given parameters $J_{\text{i}}$ and $h_{\text{i}}$ of the initial Hamiltonian $\hat{H}_{\text{i}}$ and at $t=0$ the coupling strength and the external transverse field are instantaneously changed to new values $J_{\text{f}}$ and $h_{\text{f}}$ of the final Hamiltonian $\hat{H}_{\text{f}}$ at each site of the lattice. To characterize the quenches we use the notation
\begin{align}
(J_{\text{i}};h_{\text{i}})\rightarrow(J_{\text{f}};h_{\text{f}})\;.
\end{align}
The energy change in the system due to the quench is
\begin{align}
\Delta E\equiv E_{\text{f}}-E_{\text{i},0}
\end{align}
with $E_{\text{f}}=\braket{\Psi_{\text{i},0}|\hat{H}_{\text{f}}|\Psi_{\text{i},0}}=\sum_{\lambda}E_{\text{f},\lambda}|\braket{\Psi_{\text{i},0}|\Psi_{\text{f},\lambda}}|^{2}$ the expectation value of the energy in the system after the quench. The $E_{\text{f},\lambda}$ are the eigenvalues of the final Hamiltonian $\hat{H}_{\text{f}}$ and $\ket{\Psi_{\text{f},\lambda}}$ the corresponding eigenstates.\\
More important than the energy change in the system caused by the quench is the \textit{excess energy}
\begin{align}
E_{\text{exc}}\equiv E_{\text{f}}-E_{\text{f},0}\;,
\end{align}
i.e. the energy in the system above its ground state energy after the quench. While always $E_{\text{exc}}>0$, $\Delta E$ can be positive or negative depending on the quench parameters.\\
We determine the (positive) temperature of a system in equilibrium for which the excess energy due to the quench is equal to the thermal energy above the ground state energy and compare the thermal distributions and the time averages of the distributions after the quench. The temperature attributed to a quench is called \textit{effective temperature} $T_{\text{eff}}$. Its conditional equation reads
\begin{align}
\braket{\hat{H}_{\text{f}}}_{\text{CGE}}^{T_{\text{eff}}}=E_{\text{f}}\;.
\label{Eq:conditional_equation_T_eff}
\end{align}
\begin{figure}
\includegraphics[scale=1]{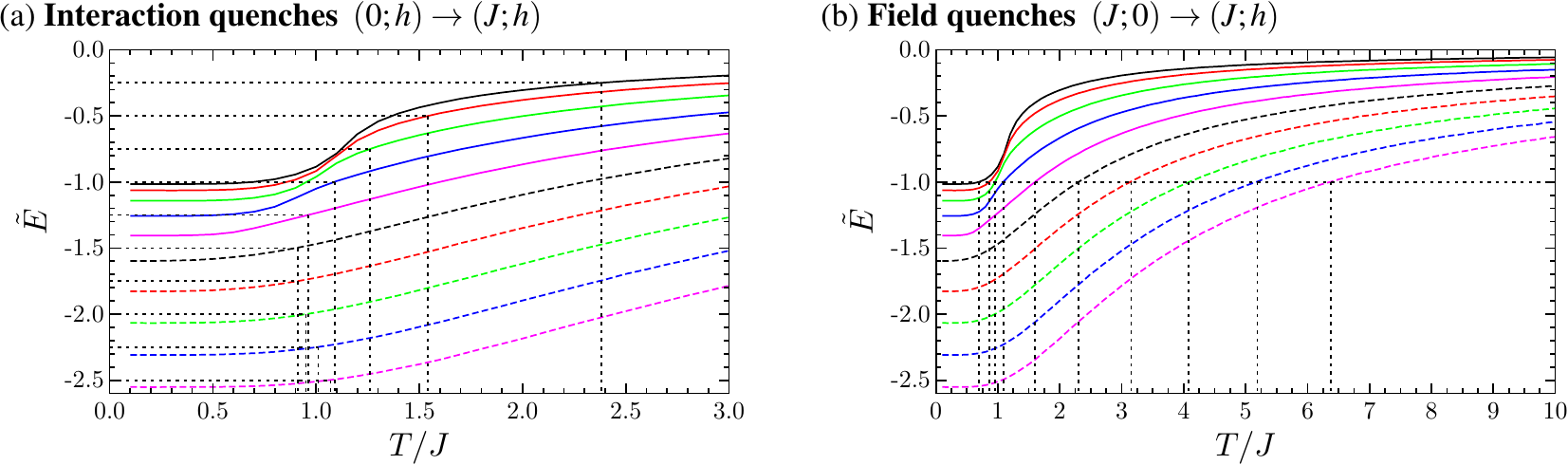}
\caption{Energy $\tilde{E}$ per site in units of $J$ as function of $T/J$ for different ratios $h/J$ for the $16\times16$ system. The graphs illustrate the determination of the effective temperature $T_{\text{eff}}$ after (a) the interaction quenches $(0;h)\rightarrow(J;h)$ and (b) the field quenches $(J;0)\rightarrow(J;h)$. The colour code is as follows: continuous lines: ${\color{black}\boldsymbol{-}}\;h/J=0.5$ / ${\color{red}\boldsymbol{-}}\;h/J=1$ / ${\color{green}\boldsymbol{-}}\;h/J=1.5$ / ${\color{blue}\boldsymbol{-}}\;h/J=2$ / ${\color{mymagenta}\boldsymbol{-}}\;h/J=2.5$, scattered lines: ${\color{black}\boldsymbol{-}}\;h/J=3$ / ${\color{red}\boldsymbol{-}}\;h/J=3.5$ / ${\color{green}\boldsymbol{-}}\;h/J=4$ / ${\color{blue}\boldsymbol{-}}\;h/J=4.5$ / ${\color{mymagenta}\boldsymbol{-}}\;h/J=5$. $\tilde{E}$ increases monotonically with $T/J$. For small $T/J$ there is a temperature interval where $\tilde{E}$ is almost constant. This temperature interval is the longer the closer $h/J$ is to the critical point $(h/J)_{\text{crit}}\approx3.044$.}
\label{fig_determination_effective_temperature}
\end{figure}
In the following we will focus on \textit{interaction quenches} $(0;h)\rightarrow(J;h)$ and \textit{field quenches} $(J;0)\rightarrow(J;h)$. These quenches just lower the ground state energy of the system, so that $\Delta E=0$ and $E_{\text{f}}=E_{\text{i},0}$. For the interaction quenches in the initial state all spins are aligned in the direction of the external transverse field:
\begin{align}
\ket{\Psi_{\text{i},0}}=\ket{\uparrow\uparrow\ldots\uparrow\uparrow}_{z}=\frac{1}{\sqrt{2^{N}}}\sum_{\mathbf{x}}\ket{\mathbf{x}}\;.
\label{Eq:GS_interaction_quench}
\end{align}
The energy of this state is $\tilde{E}_{\text{i},0}=-\frac{h}{2J}$ with the tilde denoting that the energy per site in units of the coupling constant $J$ is considered, i.e. $\tilde{E}\equiv\frac{E}{NJ}$. For the field quenches the initial state is the symmetric superposition
\begin{align}
\ket{\Psi_{\text{i},0}}=\frac{1}{\sqrt{2}}\Big\{\ket{\uparrow\uparrow\ldots\uparrow\uparrow}_{x}+\ket{\downarrow\downarrow\ldots\downarrow\downarrow}_{x}\Big\}
\label{Eq:GS_field_quench}
\end{align}
of the two fully magnetized basis states of the $\mathbf{x}$-basis with the energy $\tilde{E}_{\text{i},0}=-1$. Both the initial state of the interaction quenches and the field quenches are invariant under global spin inversion, i.e. there is no symmetry breaking of the global spin flip symmetry and thus $\braket{\Psi_{\text{i},0}|\hat{\mu}^{x}|\Psi_{\text{i},0}}=0$.\\
For the determination of the effective temperature after the quenches the thermal energy (see Figure \ref{fig_phase_diagram} (b)) of the final Hamiltonian has to be equal to the ground state energy of the system before the quench. This is illustrated in Figure \ref{fig_determination_effective_temperature}, which shows $\tilde{E}$ as a function of $T/J$ for different values of $h/J$ for the system with $L=16$. The results for $T_{\text{eff}}$ for the interaction and the field quenches are shown in Figure \ref{fig_effective_temperature} (a) and (b) respectively for different system sizes. The energy argument predicts that the end points of the interaction quenches always lie in the paramagnetic phase, i.e. starting from the completely uncorrelated state the system cannot be driven into the ferromagnetic phase switching on a coupling between the spins. There is a minimum of $T_{\text{eff}}/J$ for $h/J\approx(h/J)_{\text{crit}}$. A dependency of the effective temperature on the system size can only be observed for interaction quenches ending close to the phase transition. This is due to the increasing correlation length in the vicinity of the phase transition. Field quenches are predicted to drive the system out of the ferromagnetic phase when the external field is quenched to values larger than $h/J\approx\frac{1}{2}(h/J)_{\text{crit}}$. At the phase transition the energy isoline $\tilde{E}(h/J;T/J)=-1$, which defines the end points of the field quenches, shows a turning point, which becomes more pronounced with increasing system size. Within the ferromagnetic phase the system size has almost no effect on the effective temperature attributed to the quench. Only in the vicinity of the phase transition larger deviations between the effective temperature can be observed for different system sizes. These deviations decrease again in the paramagnetic phase.\\
\begin{figure}
\includegraphics[scale=1]{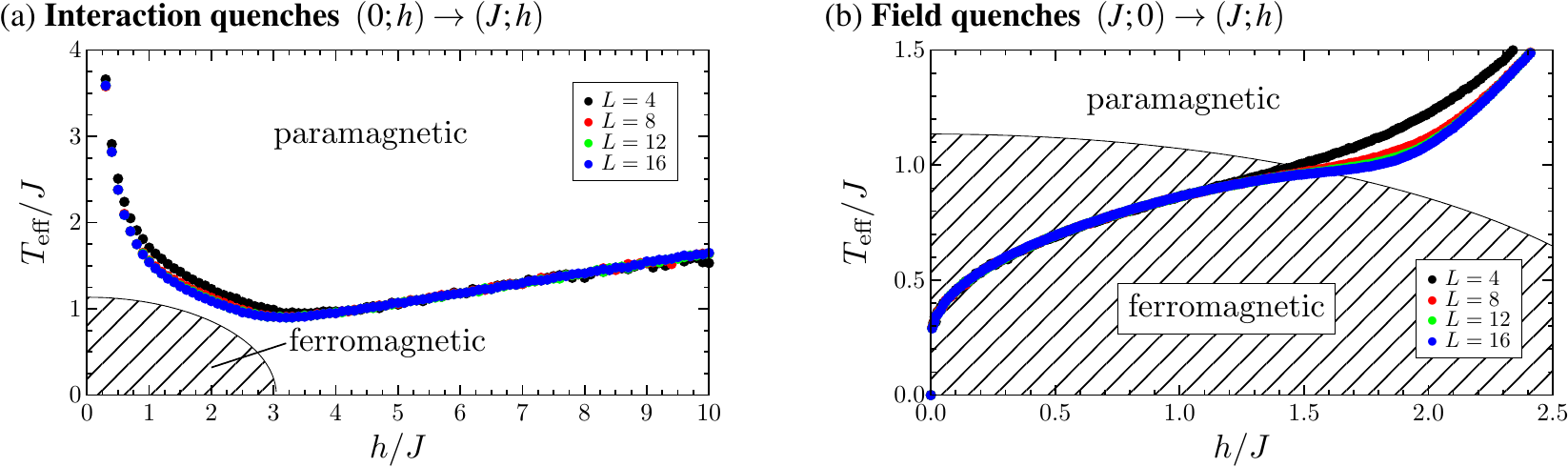}
\caption{Effective temperature $T_{\text{eff}}$ attributed to (a) interaction quenches $(0;h)\rightarrow(J;h)$ and (b) field quenches $(J;0)\rightarrow(J;h)$ for different system sizes. The shaded area represents the ferromagnetic phase for the system in the thermodynamic limit.}
\label{fig_effective_temperature}
\end{figure}
The shape of $T_{\text{eff}}/J$ as function of $h/J$ can be understood considering the effects of $J$, $h$ and $T$ onto the order in the system. In the ground state at $T=0$ for $h/J<(h/J)_{\text{crit}}$ a parallel orientation of the spins in $x$-direction is energetically preferred, while for $h/J>(h/J)_{\text{crit}}$ the orientation of the spins in $z$-direction parallel to the external transverse field is favourable. The temperature causes fluctuations of the orientation of the spins and thus disturbs the described order. The strength of this effect depends on $h/J$. For strong couplings or strong fields higher temperatures are necessary to disturb the order of the spins. This can be seen considerng the energy as a function of $T/J$ for fixed $h/J$ in Figure \ref{fig_determination_effective_temperature}. The energy increases monotonically as a function of $T/J$ with $\lim_{(T/J)\to\infty}\tilde{E}=0$. The high temperature limit follows from the symmetry of the energy eigenvalues with respect to $0$. For $T=0$ on the other hand the energy $\tilde{E}$ is close to $-1$ for small values of $h/J$, while for large ratios of $h/J$ it approaches $-\frac{h}{2J}$ from below. For low temperatures there is a temperature interval in which the energy is almost not affected by the increase of the temperature. The length of this interval depends on $h/J$. The closer $h/J$ is to $(h/J)_{\text{crit}}$, the less ordered the spins are and thus can be more easily disturbed by the temperature. As for the interaction quenches the expectation value of the energy of the system in thermal equilibrium at $T_{\text{eff}}$ has to be equal to $-\frac{h}{2J}$, this causes the minimum of $T_{\text{eff}}/J$ close to $(h/J)_{\text{crit}}$. Away from $(h/J)_{\text{crit}}$ the effective temperature of the system after the interaction quenches increases. Although for large ratios $h/J$ the energy $\tilde{E}$ at $T=0$ is only slightly smaller than the energy $-\frac{h}{2J}$ in the conditional equation for the effective temperature, due to the strong transverse field high temperatures are necessary to disturb the order of the spins. For interaction quenches with small ratios $h/J$ the same argument holds. In this case the order is caused by the coupling between the spins. For the field quenches the effective temperature is determined by the condition that $\tilde{E}=-1$. Obviously it is $T_{\text{eff}}/J=0$ for $h/J=0$. In the limit $h/J\to\infty$ on the other hand $\tilde{E}=-1$ implies $\lim_{(h/J)\to\infty}T_{\text{eff}}/J=\infty$ as the ground state energy of the system at $T=0$ is lowered almost linearly with $h$ in this case.

\subsection*{Time evolution}
We prepare the system in the ground state of the initial Hamiltonian $\hat{H}_{\text{i}}$ before the quench, i.e.
\begin{align}
\ket{\Psi(t=0)}=\ket{\Psi_{\text{i},0}}
\end{align}
with $\ket{\Psi_{\text{i},0}}$ according to equations (\ref{Eq:GS_interaction_quench}) and \,(\ref{Eq:GS_field_quench}) respectively. In general the initial state of the system is not an eigenstate of the final Hamiltonian, so that the system evolves unitarily in time according to the Schr{\"o}dinger equation
\begin{align}
\ket{\Psi(t)}=e^{-\imath\hat{H}_{\text{f}}t}\ket{\Psi(t=0)}\;.
\end{align}
In terms of the eigenbasis of $\hat{H}_{\text{f}}$ the time evolution of the expectation value of an arbitrary operator $\hat{\mathcal{O}}$ reads
\begin{align}
\braket{\hat{\mathcal{O}}}_{t}=\sum_{\lambda}|c_{\text{f},\lambda}|^{2}\mathcal{O}_{\lambda\lambda}+\sum_{\lambda\neq\lambda'}c_{\text{f},\lambda}^{*}c_{\text{f},\lambda'}e^{\imath(E_{\text{f},\lambda}-E_{\text{f},\lambda'})t}\mathcal{O}_{\lambda\lambda'}\quad\text{with}\quad c_{\text{f},\lambda}=\braket{\Psi_{\text{f},\lambda}|\Psi(t=0)}\quad,\quad\mathcal{O}_{\lambda\lambda'}=\braket{\Psi_{\text{f},\lambda}|\hat{\mathcal{O}}|\Psi_{\text{f},\lambda'}}\;.
\label{Eq:Expectation_value}
\end{align}
The initial state of the system is a pure state and its state remains a pure state under unitary time evolution. A thermal system on the other hand is described by a mixed state. For this reason time averages of the observables after the quench have to be compared to the thermal values as will be shown in the following. Considering the time-evolved expectation value in equation (\ref{Eq:Expectation_value}), one observes that the diagonal part is time-independent, while the non-diagonal contributions consist of harmonic oscillations. Averaging over time the non-diagonal part vanishes for long time intervals if there are no degenerate energy eigenvalues, so that the stationary state is determined only by the diagonal part:
\begin{align}
\lim_{\Delta t\to\infty}\frac{1}{\Delta t}\int_{t_{0}}^{t_{0}+\Delta t}dt\,\braket{\hat{\mathcal{O}}}_{t}=\sum_{\lambda}|c_{\text{f},\lambda}|^{2}\mathcal{O}_{\lambda\lambda}\;.
\end{align}
The stationary state of the system can thus be described by a mixed state in the so-called \textit{diagonal ensemble}:
\begin{align}
\braket{\hat{\mathcal{O}}}_{\text{diag}}=\text{Tr}\left[\hat{\mathcal{O}}\,\hat{\rho}_{\text{diag}}\right]\quad\text{with}\quad \hat{\rho}_{\text{diag}}=\sum_{\lambda}p_{\text{f},\lambda}\ket{\Psi_{\text{f},\lambda}}\hspace*{-0.1cm}\bra{\Psi_{\text{f},\lambda}}\quad,\quad p_{\text{f},\lambda}=|c_{\text{f},\lambda}|^{2}\;.
\end{align}
For the distributions of the (possibly degenerate) eigenvalues $\mathcal{O}_{j}$ of $\hat{\mathcal{O}}$ one has
\begin{align}
p_{\text{diag}}(\mathcal{O}_{j})=\text{Tr}\left[\delta(\mathcal{O}_{j}-\hat{\mathcal{O}})\,\hat{\rho}_{\text{diag}}\right]\;.
\end{align}
As the dimension of the Hilbert space $\mathcal{H}$ grows exponentially with the system size, the above computations in the eigenbasis of $\hat{H}_{\text{f}}$ with exact diagonalization are only possible for small systems. For this reason we use a rt-VMC method to give an accurate description of the time evolution of the 2D-TFIM for larger system sizes.

\subsection*{Real-time variational Monte Carlo}
Rt-VMC was introduced by Carleo \textit{et al.} for the Bose-Hubbard model \cite{Carleo2012,Carleo2014} and has also been successfully applied to lattice bosons and spin systems with long-range interactions \cite{Cevolani2015} as well as to strongly correlated electron systems \cite{Ido2015}. Its idea is the existence of a set of variational parameters which are sufficient to describe the physical properties of the system while their number is much smaller than the dimension of the Hilbert space. The equations of motion of the variational parameters are determined minimizing the Euclidian distance $\mathcal{D}(t)$ between the exact time evolution $\ket{\dot{\Psi}_{\text{exact}}(t)}$ of the variational state and its variational time evolution $\ket{\dot{\Psi}_{\text{var}}(t)}$, which reads in the $\mathbf{x}$-basis
\begin{align}
\mathcal{D}(t)=\sum_{\mathbf{x}}\big|\dot{\Psi}_{\text{exact}}(\mathbf{x},t)-\dot{\Psi}_{\text{var}}(\mathbf{x},t)\big|^{2}\;.
\label{Eq:Euclidian_distance}
\end{align}
A common choice for the variational state is the \textit{Jastrow ansatz} \cite{Carleo2012,Carleo2014,Cevolani2015}, which is well suited to describe the time evolution of the 2D-TFIM after interaction quenches in the paramagnetic phase. For the field quenches in the ferromagnetic phase we introduce a new ansatz, which makes use of the symmetries of the model and the high degree of order in this phase. Both ansatz functions reduce the number of parameters in the wave function of the system from a number growing exponentially with the system size to a number growing algebraically with the number of sites.

\subsubsection*{Paramagnetic phase}
In the paramagnetic phase we use the Jastrow ansatz for the variational function. This ansatz is constructed from the completely uncorrelated state. Correlations are taken into account by the \textit{Jastrow factor}. For the 2D-TFIM the Jastrow ansatz reads \cite{Cevolani2015}
\begin{align}
\ket{\Psi(t)}=\exp\Big(\sum_{\mathbf{r}}\alpha_{\mathbf{r}}(t)\hat{C}_{\mathbf{r}}^{xx}\Big)\ket{\uparrow\uparrow\dots\uparrow\uparrow}_{z}\;.
\label{Eq:Jastrow}
\end{align}
The operators
\begin{align}
\hat{C}_{\mathbf{r}}^{xx}\equiv\frac{1}{N_{\mathbf{r}}}\sum_{\mathbf{R}}\hat{\sigma}^{x}_{\mathbf{R}}\hat{\sigma}^{x}_{\mathbf{R}+\mathbf{r}}
\label{Eq:Correlation_functions}
\end{align}
measure the correlations between all spin pairs of the system with distance $\mathbf{r}$ normalized by their number $N_{\mathbf{r}}$. The sum over $\mathbf{r}$ runs over all independent directions in the lattice, whose number also determines the number of variational parameters $\alpha_{\mathbf{r}}(t)$, which is $N/8+3L/4$ for the square lattice with edge length $L$. For a given $\mathbf{r}$, the average over all dependent directions is taken (see the supplementary material). The state $\ket{\uparrow\uparrow\dots\uparrow\uparrow}_{z}$ is the completely uncorrelated state of the system, which is the exact ground state in the limit $h/J\to\infty$, i.e. which represents the completely paramagnetic state. Inserting the Jastrow ansatz from equation (\ref{Eq:Jastrow}) into equation  (\ref{Eq:Euclidian_distance}), one gets the following equations of motion for the variational parameters  \cite{Carleo2012,Carleo2014}:
\begin{align}
\sum_{\mathbf{r}'}\braket{\delta\hat{C}_{\mathbf{r}}^{xx}\delta\hat{C}_{\mathbf{r}'}^{xx}}_{t}\dot{\alpha}_{\mathbf{r}'}(t)=-\imath\braket{E^{\text{local}}_{\text{f}}(t)\delta\hat{C}_{\mathbf{r}}^{xx}}_{t}
\label{Eq:equations_of_motion_paramagnetic}
\end{align}
with $\delta\hat{\mathcal{O}}\equiv\hat{\mathcal{O}}-\braket{\hat{\mathcal{O}}}_{t}$ and $E^{\text{local}}_{\text{f}}(\mathbf{x},t)\equiv\frac{\braket{\mathbf{x}|\hat{H}_{\text{f}}|\Psi(t)}}{\braket{\mathbf{x}|\Psi(t)}}$ the local energy. The time-dependent expectation values
\begin{align}
\braket{\hat{\mathcal{O}}}_{t}\equiv\frac{\sum_{\mathbf{x}}|\Psi(\mathbf{x},t)|^{2}\mathcal{O}(\mathbf{x})}{\sum_{\mathbf{x}}|\Psi(\mathbf{x},t)|^{2}}\;.
\end{align}
have to be determined at each time step. For this we use the single spin flip quantum Monte Carlo algorithm\cite{Bishop2000} described in the supplementary material. The integration of the equations of motion is done numerically with a fourth order Runge-Kutta scheme. For the interaction quenches with $J_{\text{i}}=0$ the initial values of the variational parameters are
\begin{align}
\alpha_{\mathbf{r}}(t=0)=0\;.
\end{align}

\subsubsection*{Ferromagnetic phase}
Due to its construction from the completely paramagnetic state, the Jastrow ansatz is well suited for the paramagnetic phase, but fails to describe the time evolution of the system after field quenches in the ferromagnetic phase. For example the initial state cannot be represented by it. We thus derive in the Appendix the ansatz
\begin{align}
\ket{\Psi(t)}=\sum_{m,n}\alpha_{m,n}(t)\ket{\Psi_{m,n}}\quad\text{with}\quad\ket{\Psi_{m,n}}\equiv\frac{1}{\sqrt{N_{m,n}}}\sum_{k=1}^{N_{m,n}}\ket{\Psi_{m,n}^{k}}
\label{Eq:ansatz_ferro}
\end{align}
for the field quenches in the ferromagnetic phase. $\ket{\Psi_{m,n}}$ is the normalized symmetric superposition of all basis states with $m$ spin down and $n$ broken bonds, so-called \textit{kinks}. The ansatz separates the Hilbert space of the system into subspaces $\mathcal{H}_{m,n}$ of states with the same magnetization per site
\begin{align}
\mu^{x}_{m}=\frac{N-2m}{N}\quad\text{with}\quad m=0,1,2,\ldots,N-1,N
\label{Eq:mu}
\end{align}
and the same energy contribution of the diagonal part of the Hamiltonian per site in units of the coupling constant $J$
\begin{align}
\varepsilon^{xx}_{n}=\frac{N-n}{N}\quad\text{with}\quad n=0,4,6,\ldots,2N-6,2N-4,2N\;.
\label{Eq:varepsilon}
\end{align}
The dimension of the subspace $(m,n)$ is $N_{m,n}$. Transitions between subspaces are induced by spinflips. A spinflip increases or decreases $m$ by one and keeps $n$ untouched or changes it by $\pm2$ or $\pm4$. Thus each subspace is linked to up to $10$ other subspaces. We denote the total number of transitions between the subspaces $(m,n)$ and $(m',n')$ with $T_{m,n;m',n'}$. The $T_{m,n;m',n'}$ are symmetric with respect to $(m,n)\leftrightarrow(m',n')$. The determination of the $N_{m,n}$ and the $T_{m,n;m',n'}$ is a pure combinatorics problem. As for the 2D-TFIM there are no closed-form expressions for their values and due to the high dimensionality of the Hilbert space we determine them with rare event sampling (RES) \cite{Hartmann2002} described in the supplementary material. As the $N_{m,n}$ and $T_{m,n;m',n'}$ are independent of the quench parameters, they have to be determined only once for each system size. The possible values of $n$ depend on $m$ in a non-trivial way. In case of the 2D-TFIM not just the maximal number of kinks is a function of $m$ like in one dimension, but also their minimal number. The PBC have to be taken into account, too. In leading order the number of possible values of $n$ grows linearly with $m$. As the number of possible values of $m$ grows linearly with the system size $N$, the number of variational parameters thus grows in leading order with $N^{2}$. Due to the symmetry of the variational parameters with respect to $m\leftrightarrow N-m$ and the conservation of this symmetry under time evolution, we can reduce the number of independent variational parameters using $\alpha_{m,n}(t)=\alpha_{N-m,n}(t)$. For the system sizes we studied the numbers of the variational parameters are listed in Table \ref{Tab:number_of_variational_parameters}.\\
\begin{table}
\renewcommand{\arraystretch}{1.25}
\centering
\begin{tabular}{>{\centering}m{2.5cm}||>{\centering}m{1.0cm}|>{\centering}m{1.0cm}|>{\centering}m{1.0cm}|>{\centering}m{1.0cm}}
$L$ & $4$ & $8$ & $12$ & $16$\tabularnewline
\hline
number of $\alpha_{m,n}$ & $45$ & $848$ & $4551$ & $14834$
\end{tabular}
\caption{Number of variational parameters for the ansatz function in equation (\ref{Eq:ansatz_ferro}) for the field quenches as function of the edge length $L$ of the square lattice. In leading order the growth is proportional to $N^{2}$.
\label{Tab:number_of_variational_parameters}}
\end{table}
The equations of motion of the variational parameters are derived in the Appendix:
\begin{align}
\imath\dot{\alpha}_{m,n}(t)=-J\,(N-n)\,\alpha_{m,n}(t)-\frac{h}{2}\sum_{m',n'}t_{m',n';m,n}\,\alpha_{m',n'}(t)\quad\text{with}\quad t_{m',n';m,n}\equiv\frac{T_{m',n';m,n}}{\sqrt{N_{m',n'}\,N_{m,n}}}\;.
\label{Eq:equations_of_motion_ferromagnetic}
\end{align}
The sum over $m'$ and $n'$ runs over all subspaces that can be reached from any basis state of the subspace $(m,n)$ by flipping one single spin. The equations of motion of the variational parameters are thus a system of coupled linear differential equations of first order with constant (time-independent) coefficients, which are known from RES. As each subspace is linked to only up to $10$ other subspaces, the system is sparse. We solve it with a fourth order Runge-Kutta scheme. The initial values of the variational parameters are
\begin{align}
\alpha_{m,n}(t=0)=
\begin{cases}
\newline \frac{1}{\sqrt{2}} & \newline \text{if } (m,n)=(0,0)\text{ or } (N,0)\\
\newline 0 & \newline \text{else}
\end{cases}
\;.
\end{align}
For $t>0$ the $\alpha_{m,n}(t)$ are in general complex.

\subsection*{Observables}
As observables we consider the rescaled modulus of the magnetization according to equation (\ref{Eq:mu_tilde}) and the correlation function between two spins at distance $\mathbf{r}$ according to equation (\ref{Eq:Correlation_functions}), which reads for nearest neighbours
\begin{align}
\hat{C}_{\text{nn}}^{xx}=\frac{1}{2N}\sum_{<\mathbf{R},\mathbf{R}'>}\hat{\sigma}^{x}_{\mathbf{R}}\hat{\sigma}^{x}_{\mathbf{R}'}\;.
\end{align}
For the interaction quenches the expectation values of the observables and their distributions are computed in the course of the single spin flip quantum Monte Carlo algorithm for the coefficients of the equations of motion at each time step, while for the field quenches there is a direct functional relationship to the variational parameters. For the modulus of the magnetization this relationship reads
\begin{align}
\braket{|\hat{\mu}^{x}|}_{t}=\sum_{m,n}|\alpha_{m,n}(t)|^{2}\cdot|\mu^{x}_{m}|
\end{align}
with the eigenvalues $\mu^{x}_{m}$ of $\hat{\mu}^{x}$ according to equation (\ref{Eq:mu}) and for the correlation function between nearest neighbours it is
\begin{align}
\braket{\hat{C}_{\text{nn}}^{xx}}_{t}=\sum_{m,n}|\alpha_{m,n}(t)|^{2}\cdot\varepsilon^{xx}_{n}
\end{align}
with the eigenvalues $\varepsilon^{xx}_{n}$ of $\hat{C}_{\text{nn}}^{xx}$ according to equation (\ref{Eq:varepsilon}). The distributions of $\mu^{x}_{m}$ and $\varepsilon^{xx}_{n}$ at time $t$ are given by
\begin{align}
p_{t}(\mu^{x}_{m})=\sum_{n}|\alpha_{m,n}(t)|^{2}
\end{align}
and
\begin{align}
p_{t}(\varepsilon^{xx}_{n})=\sum_{m}|\alpha_{m,n}(t)|^{2}\;.
\end{align}
The correlation function between spins that are not nearest neighbours can be computed according to
\begin{align}
\braket{\hat{C}_{\mathbf{r}}^{xx}}_{t}=\sum_{m,n}|\alpha_{m,n}(t)|^{2}\cdot C_{\mathbf{r}}^{xx}(m,n)\quad\text{with}\quad C_{\mathbf{r}}^{xx}(m,n)\equiv\frac{1}{N_{\mathbf{r}}}\sum_{\mathbf{R}}\sum_{k}\underbrace{\braket{\Psi_{m,n}^{k}|\hat{\sigma}^{x}_{\mathbf{R}}\hat{\sigma}^{x}_{\mathbf{R}+\mathbf{r}}|\Psi_{m,n}^{k}}}_{\pm1}\;.
\end{align}
The $C_{\mathbf{r}}^{xx}(m,n)$ also have to be determined with RES.\\
To decide whether the system thermalizes or not the expectation values of the observables as well as their distributions in the stationary state have to be compared to their counterparts for the thermal system. As the exact computation in the diagonal ensemble would require the knowledge of the full spectrum of the Hamiltonian after the quench, we use time averages to approximate the expectation values and distributions in the stationary state according to
\begin{align}
\overline{\braket{\hat{\mathcal{O}}}_{t}}=\frac{1}{\Delta t}\int_{t_{0}}^{t_{0}+\Delta t}dt\,\text{Tr}\left[\hat{\mathcal{O}}\,\hat{\rho}(t)\right]
\end{align}
and
\begin{align}
\overline{p_{t}(\mathcal{O}_{j})}=\frac{1}{\Delta t}\int_{t_{0}}^{t_{0}+\Delta t}dt\,\text{Tr}\left[\delta(\mathcal{O}_{j}-\hat{\mathcal{O}})\,\hat{\rho}(t)\right]
\end{align}
with $\hat{\rho}(t)=\ket{\Psi(t)}\hspace*{-0.1cm}\bra{\Psi(t)}$ and $\mathcal{O}_{j}$ the (possibly degenerate) eigenvalues of $\hat{\mathcal{O}}$. We computed the time averages for different interval lengths $\Delta t$ and different $t_{0}$ in the range of times that we can simulate ($t_{0}+\Delta t<25$ for the $16\times16$ system, longer times for smaller systems) and found that the results are stable with respect to our tests.\\
The time averages of the observables are compared to the thermal expectation values of the system in equilibrium at the effective temperature $T_{\text{eff}}$ attributed to the quench, which are given by
\begin{align}
\braket{\hat{\mathcal{O}}}_{\text{CGE}}^{T_{\text{eff}}}=\frac{1}{Z_\text{CGE}^{T_{\text{eff}}}}\text{Tr}\left[\hat{\mathcal{O}}e^{-\hat{H}/T_{\text{eff}}}\right]\quad\text{with}\quad Z_\text{CGE}^{T_{\text{eff}}}=\text{Tr}\left[e^{-\hat{H}/T_{\text{eff}}}\right]\;.
\end{align}
The thermal distribution of the eigenvalues $\mathcal{O}_{j}$ of $\hat{\mathcal{O}}$ reads
\begin{align}
p_{\text{CGE}}^{T_{\text{eff}}}(\mathcal{O}_{j})=\frac{1}{Z_\text{CGE}}\text{Tr}\left[\delta(\mathcal{O}_{j}-\hat{\mathcal{O}})e^{-\hat{H}/T_{\text{eff}}}\right]\;.
\end{align}
The expectation values and distributions for the system in thermal equilibrium are computed with a cluster Monte Carlo algorithm in continuous imaginary time\cite{Rieger1999}.

\section*{Results and discussion}
\begin{figure}
\includegraphics[scale=1]{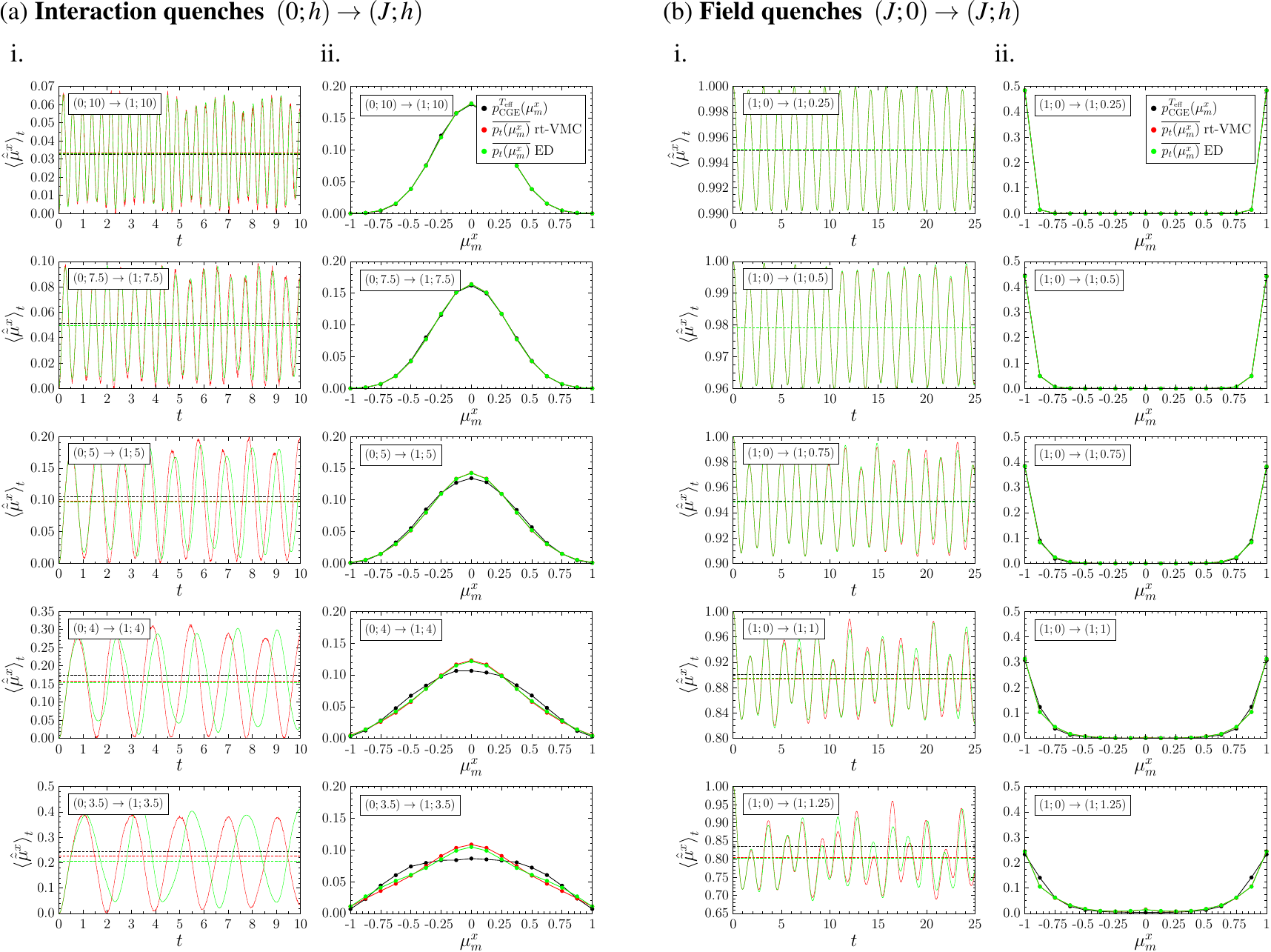}
\caption{Comparison between thermal values (${\color{black}\bullet}$), results of the rt-VMC algorithm (${\color{red}\bullet}$) and results of exact diagonalization (${\color{green}\bullet}$) for the rescaled modulus of the magnetization after (a) interaction quenches and (b) field quenches for the $4\times4$ system. (a) i. and (b) i. contain a comparison of the exact and the rt-VMC time evolution, (a) ii. and (b) ii. a comparison between the thermal distribution and the time-averaged distributions of the rt-VMC and the exact computations with $\Delta t=100$. The thermal values have been computed for the system in equilibrium at the temperature $T_{\text{eff}}$ attributed to the quench. The distance of the end point of the quenches from the phase transition is reduced from top to bottom.}
\label{fig_4_times_4}
\end{figure}

\begin{figure}
\includegraphics[scale=1]{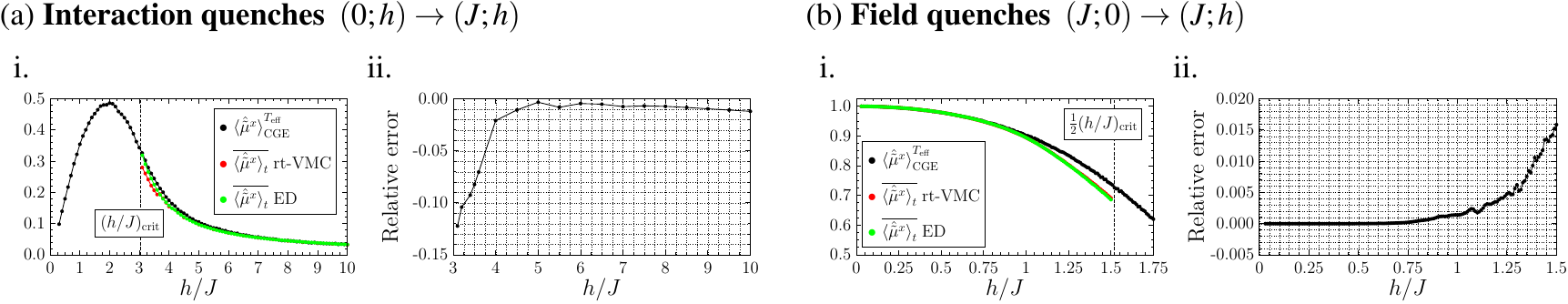}
\caption{Comparison between results of exact diagonalization (${\color{green}\bullet}$) and rt-VMC (${\color{red}\bullet}$) for the rescaled modulus of the magnetization after (a) interaction quenches and (b) field quenches for the $4\times4$ system to thermal values for the system in equilibrium at the temperature $T_{\text{eff}}$ (${\color{black}\bullet}$). The graphs (a) ii. and (b) ii. show the relative error of the rt-VMC time averages compared to the time averages of the exact time evolution.}
\label{fig_comparison}
\end{figure}

Before we apply the rt-VMC algorithm to large system sizes, we consider a system of size $4\times4$, whose time evolution can also be computed with numerical integration of the Schr{\"o}dinger equation, and compare the rt-VMC results to the exact results to benchmark the algorithm. We do this exemplarily for the rescaled modulus of the magnetization. Results are presented in Figure \ref{fig_4_times_4} for (a) interaction quenches and (b) field quenches. (a) i. and (b) i. contain a comparison between the exact time evolution (green) and the rt-VMC time evolution (red). Time averages are represented by the dashed lines of the respective colours, while the black dashed line is the thermal expectation value for the system in equilibrium at the temperature attributed to the quench. In (a) ii. and (b) ii. we compare the time-averaged distributions of $\mu^{x}_{m}$ of the exact diagonalization (green) and the rt-VMC (red) to the thermal distribution (black). We observe that after the interaction quenches the shape of the curves of the time evolution is close to harmonic oscillations with time-dependent variations of the amplitude. As long as $h/J\gg(h/J)_{\text{crit}}$, the rt-VMC algorithm with the Jastrow ansatz allows a good description of the time evolution. There are differences between the frequencies and between the amplitudes, which increase for stronger interaction quenches when the system is driven closer to its phase transition.
Despite of these deviations the time averages as well as the time-averaged distributions of the exact and the rt-VMC time evolution still show a very good agreement except for the quench $(0;3.5)\rightarrow(1;3.5)$, which drives the system close to its phase transition. Deviations from the thermal values on the other hand can already be observed beginning from the quench $(0;5)\rightarrow(1;5)$. For the field quenches the rt-VMC results for the time evolution after the quench show an even better agreement with the exact time evolution. The frequency of the oscillations is reproduced with high accuracy even for strong quenches. With increasing quench strength small deviations of the amplitude of the oscillations can be observed, but the time averages and the time-averaged distributions of the rt-VMC and the corresponding results of the exact time evolution coincide for all the quenches we studied. As in case of the interaction quenches there are increasing deviations between the time averages after the quench and the results for the thermal system in equilibrium at $T_{\text{eff}}$ with increasing quench strength. In order to quantify the deviations between the thermal expectation values and the time averages after the quench as a function of the quench parameters we compare in Figure \ref{fig_comparison} (a) i. and (b) i. the thermal expectation values for the system in equilibrium at the temperature $T_{\text{eff}}$ attributed to the quench (black) and the time-averaged values of the exact time evolution (green) and of the rt-VMC time evolution (red). The relative error between the time-averaged expectation values of the exact and the rt-VMC time evolution is shown in (a) ii. and (b) ii.. For the interaction quenches we observe an excellent agreement while $h/J\gtrsim4$ with a deviation of less than $2\,\%$. For the field quenches we find that beginning from $h/J\approx0.75$ deviations increase, but even for $h/J=1.5$ they do not exceed $1.5\,\%$. For both quench protocols our rt-VMC method thus allows us to compute the time averages of the observables as well as their distributions with high accuracy for a wide range of ratios $h/J$. We may derive two main results from our studies of the $4\times4$ system: First we observe that for this small system size significant deviations between time-averaged results and the thermal results exist when the system is quenched close to its phase transition. Second there is a very good agreement between the time-averaged values of the exact time evolution and the rt-VMC time evolution for a wide range of ratios $h/J$. This agreement does not just concern the time averages of the observables, but also the underlying distributions. Only for strong quenches deviations between the time averages of the exact time evolution and the rt-VMC time evolution can be observed, but these deviations are much smaller than the deviations from the values of the system in thermal equilibrium.\\
\begin{figure}
\includegraphics[scale=1]{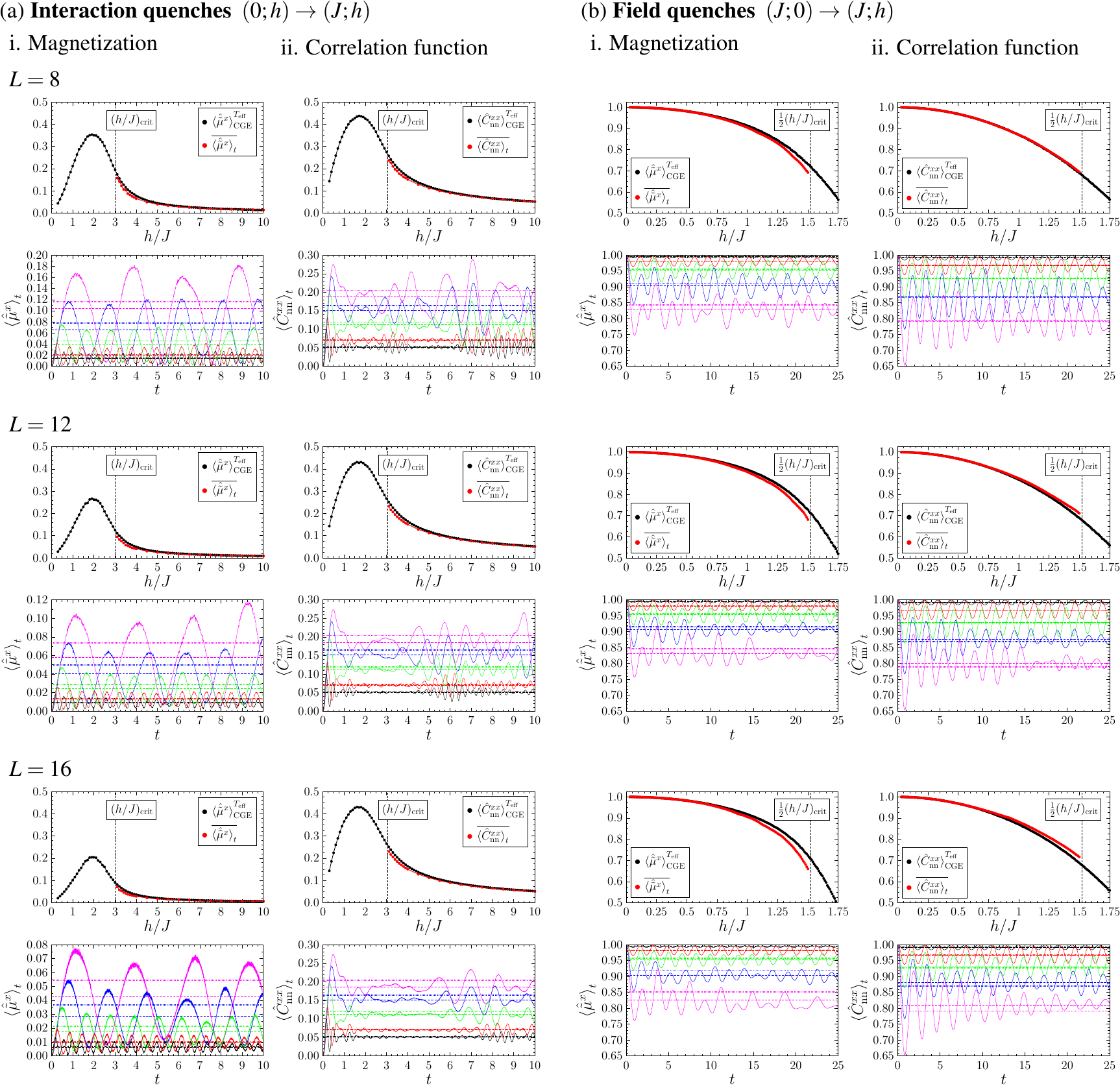}
\caption{Results for i. the rescaled modulus of the magnetization and ii. the correlation function between nearest neighbours after (a) interaction quenches and (b) after field quenches for the system sizes $L=8$, $12$ and $16$. The graphs in the first line for each system size show the time evolution of the observables (rescaled modulus of the magnetization and correlation function between nearest neighbours) after the quenches for different quench parameters (continuous lines), their time averages (dashed lines) and the thermal expectation values for the system in equilibrium at $T_{\text{eff}}$ according to the CGE (dotted lines). To characterize the quenches we use the notation $(J_{\text{i}};h_{\text{i}})\rightarrow(J_{\text{f}};h_{\text{f}})$. For the interaction quenches the colour code is as follows: ${\color{black}\boldsymbol{-}}\;(0;10)\rightarrow(1;10)$ / ${\color{red}\boldsymbol{-}}\;(0;7.5)\rightarrow(1;7.5)$ / ${\color{green}\boldsymbol{-}}\;(0;5)\rightarrow(1;5)$ / ${\color{blue}\boldsymbol{-}}\;(0;4)\rightarrow(1;4)$ / ${\color{mymagenta}\boldsymbol{-}}\;(0;3.5)\rightarrow(1;3.5)$; and for the field quenches: ${\color{black}\boldsymbol{-}}\;(1;0)\rightarrow(1;0.25)$ / ${\color{red}\boldsymbol{-}}\;(1;0)\rightarrow(1;0.5)$ / ${\color{green}\boldsymbol{-}}\;(1;0)\rightarrow(1;0.75)$ / ${\color{blue}\boldsymbol{-}}\;(1;0)\rightarrow(1;1)$ / ${\color{mymagenta}\boldsymbol{-}}\;(1;0)\rightarrow(1;1.25)$. In the graphs in the second line for each system size we compare the thermal expectation values of the observables (${\color{black}\bullet}$) to their time averages after the quenches (${\color{red}\bullet}$) as a function of $h/J$ after the quench. For the interaction quenches the system does not leave the paramagnetic phase, while field quenches will drive the system from the ferromagnetic into the paramagnetic phase for $h/J\gtrsim\frac{1}{2}(h/J)_{\text{crit}}$. We observe continuously increasing deviations between the thermal expectation values and the time averages after the quenches the closer the system is quenched to its phase transition.}
\label{fig_results}
\end{figure}
We now apply the rt-VMC method to larger system sizes. To make predictions for the system in the thermodynamic limit, we computed results for the rescaled modulus of the magnetization and the correlation function between nearest neighbours after interaction quenches and after field quenches for systems of size $L=8$, $12$ and $16$. Figure \ref{fig_results} shows our results for the time evolution of the observables for the different quench protocols we considered as well as a comparison between their time averages and the thermal expectation values for the system in equilibrium at $T_{\text{eff}}$ as a function of $h/J$ after the quench. For the interaction quenches we confine us to values $h/J>(h/J)_{\text{crit}}$, as for smaller ratios $h/J$ the accuracy of the rt-VMC with the Jastrow ansatz decreases, while for the field quenches we study quenches with $h/J<\frac{1}{2}(h/J)_{\text{crit}}$ as for stronger quenches the system is predicted to be driven from the ferromagnetic to the paramagnetic phase. We observe that after the quenches the observables quickly approach a stationary value and in the following oscillate around it. This stationary value is in good approximation constant for the simulated time intervals. For small quenches both the amplitude and the frequency of the oscillations are almost constant. For larger quenches modulations of the amplitudes occur. Considering the time averages and the thermal expectation values as function of $h/J$ after the quench we observe that the relative shape of the curves is preserved when the system size is increased. There is a good agreement between the time averages and the thermal expectation values for small quenches with continuously increasing deviations for stronger quenches. While for the interaction quenches the observed deviations are small, in case of the field quenches the deviations become significant when the system is quenched closer to the phase transition. Comparing the deviations for strong field quenches one observes that they do not decrease with the system size, so that we assume them not to be caused by finite size effects.\\
Up to this point we have only considered the time-dependent expectation values of the observables and compared their time averages to the thermal expectation values. To decide whether the stationary state of the system after the quenches can be described by the CGE not only the expectation values of the observables have to coincide, but also their distributions. The study of the distributions is thus especially important for the interaction quenches and small field quenches, for which we have found a good agreement between the thermal and the time-averaged expectation values of the rescaled modulus of the magnetization as well as of the correlation function between nearest neighbours. [The strong variations in the distributions of the correlation function between nearest neighbours for $\varepsilon^{xx}_{n}$ close to $+1$ are not due to deficiencies of the algorithm, but are intrinsic to the system. Values of $\varepsilon^{xx}_{n}$ close to $+1$ correspond to small kink numbers $n$. The possible values of $n$ depend on the number $m$ of spin down. In the two-dimensional model the smallest possible value of $n$ for a given $m$ increases with $m$. For this reason small values of $n$ are linked to small values of $m$. For small values of $m$ the configurations with the highest possible number of kinks are much more likely than configurations with the smallest possible number of kinks. This causes the strong variations in the distributions of $\varepsilon^{xx}_{n}$ close to $+1$.] As the number of possible values of $\mu^{x}_{m}$ and $\varepsilon^{xx}_{n}$ grows linearly with the system size, the resolution of the distributions becomes higher with increasing system size. For this reason we show in Figure \ref{fig_distributions} the distributions of $\mu^{x}_{m}$ and $\varepsilon^{xx}_{n}$ for the $16\times16$ system, i.e. the largest system size we can simulate. The quench protocols for the interaction and the field quenches are the same as in Figure \ref{fig_results}. The distance of the end point of the quenches from the phase
\begin{figure}
\includegraphics[scale=1]{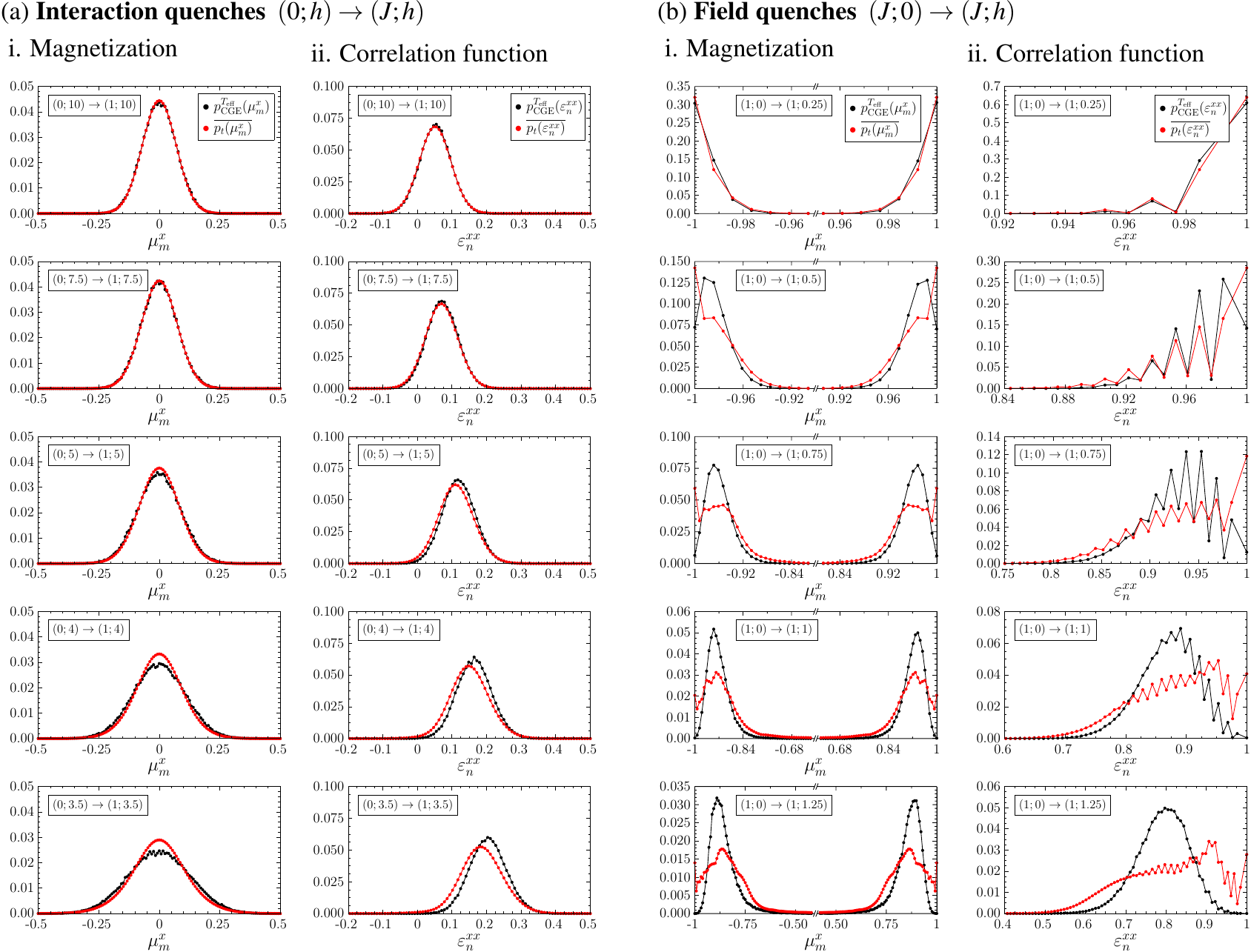}
\caption{Comparison between the thermal (${\color{black}\bullet}$) and the time-averaged distributions (${\color{red}\bullet}$) of i. $\mu^{x}_{m}$ and ii. $\varepsilon^{xx}_{n}$ after (a) interaction quenches and (b) field quenches for a system of size $L=16$. The quench protocols are the same as in Figure \ref{fig_results}. The quench strength is increased from top to bottom. Especially in case of the field quenches we observe clear deviations between the thermal and the time-averaged distributions already for small quenches, for which the thermal expectation values and the time-averages still agree.}
\label{fig_distributions}
\end{figure}
transition is reduced from top to bottom, i.e. for the interaction quenches the ratio $h/J$ after the quench is decreased, while for the field quenches it is increased. For small interaction quenches we observe a very good agreement between the thermal and the time-averaged distributions. Deviations increase when the system is quenched closer to the phase transition, but the shape of the thermal curves is well reproduced by the time averages after the quenches. Comparing the distributions for the $16\times16$ system to those of the $4\times4$ system in Figure \ref{fig_4_times_4} we observe that the agreement is better than for the smaller system. For the field quenches we find strong deviations between the thermal distributions and the time-averaged distributions. In the distributions deviations can already be observed for quench protocols for which the time averages still agree with the thermal expectation values. These deviations could not be seen in the distributions of the $4\times4$ system in Figure \ref{fig_4_times_4} due to the lower resolution of the distributions caused by the small system size. In general we observe that the time-averaged distributions after the field quenches are wider than the thermal distributions. The positions of their maxima are almost the same as for the thermal distributions, but the maxima are less pronounced. In addition the time-averaged distributions after the quenches show an increased probability to find the system in the fully ordered state ($\mu^{x}_{m}=\pm1$ or $\varepsilon^{xx}_{n}=+1$ respectively) compared to the thermal system. As these states are the initial state of the system this means that the system does not lose the memory of its initial state which contradicts to thermalization. We can thus state that for the system sizes and time scales that we can simulate the system does not thermalize after field quenches.\\
Up to this point we have only given a qualitative discussion of the distributions of the observables. In order to compare the deviations as a function of the system size and make predictions for the system in the thermodynamic limit we introduce the measure for the deviations between the thermal and the time-averaged distributions
\begin{align}
\Delta(\hat{\mathcal{O}})=\frac{1}{N(\mathcal{O}_{j})}\sum_{j}\frac{|\overline{p_{t}(\mathcal{O}_{j})}-p_{\text{CGE}}^{T_{\text{eff}}}(\mathcal{O}_{j})|}{p_{\text{CGE,max}}^{T_{\text{eff}}}}\quad\text{with}\quad p_{\text{CGE,max}}^{T_{\text{eff}}}=\max\left\{p_{\text{CGE}}^{T_{\text{eff}}}(\mathcal{O}_{j})\right\}\;.
\label{Eq:difference_distributions}
\end{align}
\begin{figure}
\includegraphics[scale=1]{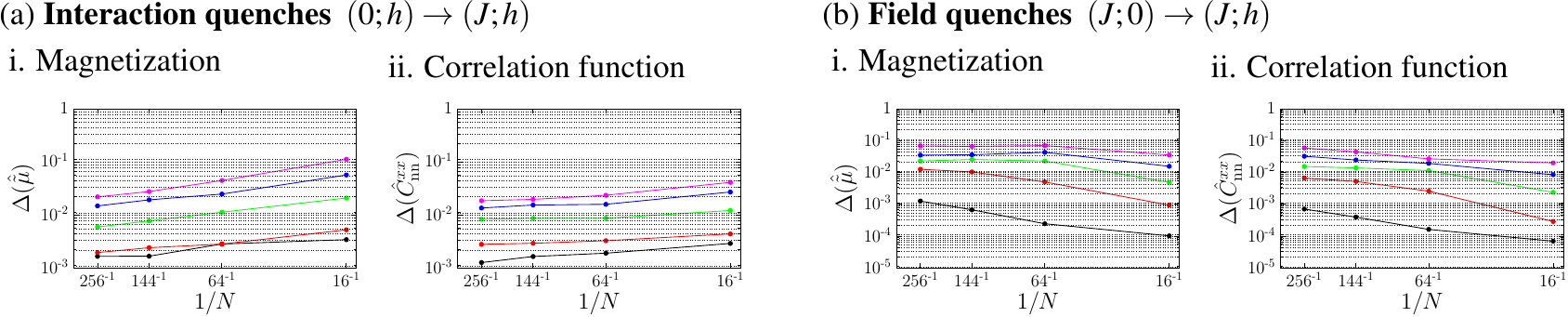}
\caption{Deviations between the thermal distributions and the time-averaged distributions for i. the rescaled modulus of the magnetization and ii. the correlation function between nearest neighbours after (a) interaction quenches and (b) field quenches as function of the inverse system size for the system sizes $L=4$, $8$, $12$ and $16$. The deviations $\Delta(\hat{\tilde{\mu}})$ and $\Delta(\hat{C}_{\text{nn}}^{xx})$ respectively between the distributions have been computed according to equation (\ref{Eq:difference_distributions}). For the interaction quenches the colour code reads: ${\color{black}\bullet}\;(0;10)\rightarrow(1;10)$ / ${\color{red}\bullet}\;(0;7.5)\rightarrow(1;7.5)$ / ${\color{green}\bullet}\;(0;5)\rightarrow(1;5)$ / ${\color{blue}\bullet}\;(0;4)\rightarrow(1;4)$ / ${\color{mymagenta}\bullet}\;(0;3.5)\rightarrow(1;3.5)$; and for the field quenches: ${\color{black}\bullet}\;(1;0)\rightarrow(1;0.25)$ / ${\color{red}\bullet}\;(1;0)\rightarrow(1;0.5)$ / ${\color{green}\bullet}\;(1;0)\rightarrow(1;0.75)$ / ${\color{blue}\bullet}\;(1;0)\rightarrow(1;1)$ / ${\color{mymagenta}\bullet}\;(1;0)\rightarrow(1;1.25)$. One observes that for the interaction quenches the differences between the distributions decrease with increasing system size, while for the field quenches they increase or remain almost constant.}
\label{fig_differences_distributions}
\end{figure}
$N(\mathcal{O}_{j})$ is the number of different eigenvalues of $\hat{\mathcal{O}}$. The normalization is with respect to the maximum of the thermal distribution for the considered quench protocol and system size. We computed $\Delta(\hat{\tilde{\mu}}^{x})$ and $\Delta(\hat{C}_{\text{nn}}^{xx})$ for the system sizes $L=4$, $8$, $12$ and $16$ for the described quench protocols and did a finite size scaling to conclude to the system in the thermodynamic limit. Figure \ref{fig_differences_distributions} shows i. $\Delta(\hat{\tilde{\mu}}^{x})$ and ii. $\Delta(\hat{C}_{\text{nn}}^{xx})$ after (a) the interaction and (b) the field quenches for the different quench protocols. The results are plotted as a function of the inverse system size $1/N$. We find that for the interaction quenches the deviations between the thermal and the time-averaged distribution decrease with increasing system size. We thus conclude that the observed deviations of the expectation values and the distributions between the thermal and the time-averaged system will further decrease with increasing system size and that the system will thermalize in the thermodynamic limit. Before we make a statement on the results of the finite size scaling for the field quenches we discuss the effect of the spontaneous symmetry breaking in the ferromagnetic phase for the system in the thermodynamic limit. For the finite system sizes considered in our numerical studies the initial state of the field quenches is the symmetric superposition of the two fully magnetized states according to equation (\ref{Eq:GS_field_quench}). In the thermodynamic limit on the other hand $\braket{\Psi_{\text{i},0}|\hat{\mu}^{x}|\Psi_{\text{i},0}}=\pm1$ implies $\ket{\Psi_{\text{i},0}}=\ket{\uparrow\uparrow\ldots\uparrow\uparrow}_{x}$ and $\ket{\Psi_{\text{i},0}}=\ket{\downarrow\downarrow\ldots\downarrow\downarrow}_{x}$ respectively. We compute the time evolution of the expectation value of an arbitrary operator $\hat{\mathcal{O}}$ starting from the symmetric superposition and compare it to the time evolution starting from $\ket{\uparrow\uparrow\ldots\uparrow\uparrow}_{x}$:
\begin{align}
\begin{split}
\braket{\hat{\mathcal{O}}}_{t}=\frac{1}{2}\Big\{&\leftidx{_{x}\hspace*{-0.05cm}}{\braket{\uparrow\uparrow\ldots\uparrow\uparrow|e^{\imath\hat{H}t}\hat{\mathcal{O}}e^{-\imath\hat{H}t}|\uparrow\uparrow\ldots\uparrow\uparrow}}_{x}+\leftidx{_{x}\hspace*{-0.05cm}}{\braket{\downarrow\downarrow\ldots\downarrow\downarrow|e^{\imath\hat{H}t}\hat{\mathcal{O}}e^{-\imath\hat{H}t}|\downarrow\downarrow\ldots\downarrow\downarrow}}_{x}\\
&+\leftidx{_{x}\hspace*{-0.05cm}}{\braket{\uparrow\uparrow\ldots\uparrow\uparrow|e^{\imath\hat{H}t}\hat{\mathcal{O}}e^{-\imath\hat{H}t}|\downarrow\downarrow\ldots\downarrow\downarrow}}_{x}+\leftidx{_{x}\hspace*{-0.05cm}}{\braket{\downarrow\downarrow\ldots\downarrow\downarrow|e^{\imath\hat{H}t}\hat{\mathcal{O}}e^{-\imath\hat{H}t}|\uparrow\uparrow\ldots\uparrow\uparrow}}_{x}\Big\}\;.
\end{split}
\end{align}
For the rescaled modulus of the magnetization $\hat{\tilde{\mu}}^{x}$ and the correlation function $\hat{C}_{\mathbf{r}}^{xx}$ the first and the second expectation value in the sum give the same result. Their sum thus just corresponds to the time evolution starting from $\ket{\uparrow\uparrow\ldots\uparrow\uparrow}_{x}$. Applying a series expansion of the time evolution operator one can easily show that the two remaining matrix elements vanish in the thermodynamic limit, so that the time-evolved expectation values of $\hat{\tilde{\mu}}^{x}$ and $\hat{C}_{\mathbf{r}}^{xx}$ are independent of the initial state for the system in the thermodynamic limit. Simulations for finite system sizes indeed show that the difference between the time evolutions starting from the two different initial states decrease with increasing system size. Our results for the field quenches can thus be used to conclude to the system in the thermodynamic limit. The finite size scaling in Figure \ref{fig_differences_distributions} (b) shows that both $\Delta(\hat{\tilde{\mu}}^{x})$ and $\Delta(\hat{C}_{\text{nn}}^{xx})$ increase with the system size or at least do not decrease. We interpret this as indication that the system will not completely thermalize in the thermodynamic limit either.\\
\begin{figure}
\includegraphics[scale=1]{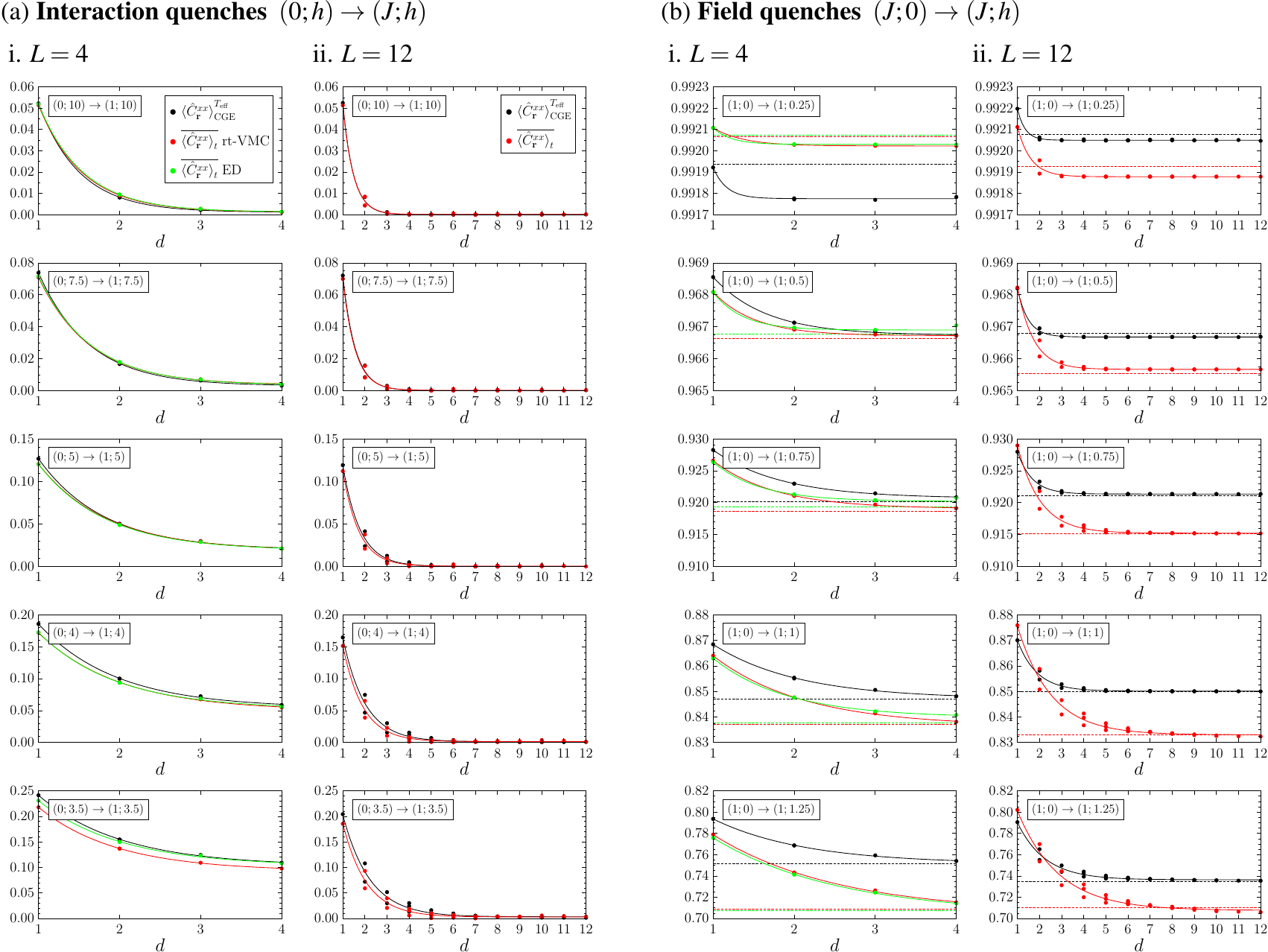}
\caption{Correlation function between spins at arbitrary distance for i. the $4\times4$ and ii. the $12\times12$ system after (a) interaction quenches and (b) field quenches. The distance $d$ is measured in the Manhattan metric. We compare the thermal expectation values for the system in equilibrium at temperature $T_{\text{eff}}$ (${\color{black}\bullet}$) to the time-averaged values after the quenches computed with rt-VMC (${\color{red}\bullet}$). For the $4\times4$ system we additionally include the time-averaged values after the quenches computed with exact diagonalization (${\color{green}\bullet}$). The continuous curves in the graphs for the interaction quenches are least-square fits according to an exponential decay $a\cdot e^{bd}$ as predicted by analytic expressions. For the field quenches the continuous lines serve as a guide to the eye and the dotted lines represent the square of the rescaled modulus of the magnetization.}
\label{fig_long_range_correlations}
\end{figure}
As the last point of our studies we consider correlation functions between spins which are not nearest neighbours. Due to the long-range order in the ferromagnetic phase the correlation function of two spins will not vanish at large distances but reach a constant non-vanishing value given by the square of the expectation value of the rescaled modulus of the magnetization. In the paramagnetic phase on the other hand there is no long-range order and the Hamiltonian is gapped, i.e. there is an energy gap between the ground state and the first excited state. It has been shown that in the ground state ($T=0$) of gapped quantum systems correlations decay exponentially with the distance \cite{Hastings2004,Nachtergaele2006}. The same holds for thermal states in the paramagnetic phase \cite{Kliesch2014}. In contrast to this there are no analytic expressions for the decay of the correlations in phases with long-range order, i.e. the ferromagnetic phase in case of the 2D-TFIM. We compute the long-range correlations for the $4\times4$ and the $12\times12$ system. The constraint to the $12\times12$ system is due to the computational effort in the computation of the $C_{\mathbf{r}}^{xx}(m,n)$. The rt-VMC results for the $4\times4$ system are compared to exact results. Figure \ref{fig_long_range_correlations} shows the results for the long-range correlations for i. the $4\times4$ system and ii. the $12\times12$ system after (a) interaction quenches and (b) after field quenches. The distance $d$ between two sites is measured in the Manhattan metric, i.e. if $\mathbf{r}=\mathbf{R}-\mathbf{R}'$ defines the relative position of the sites we have $d=\sum_{i=1}^{D}|R_{i}-R_{i}'|$ with $D$ the dimensionality (here $D=2$). Thus for a given distance $d$ there may be several $\mathbf{r}$. The strength of the correlation between two sites depends on the number of shortest paths between them, which depends on $\mathbf{r}$. We compare the time-averaged rt-VMC results after the quenches (red) to the correlation functions for the system in equilibrium at the temperature $T_{\text{eff}}$ attributed to the quench (black). For the $4\times4$ system we additionally show results of the exact time evolution (green). For the interaction quenches we do a least-square fit with $a\cdot e^{bd}$. We observe that the decay of the correlations in the thermal system as well as in the quenched system is well described by the fit curves in agreement with the analytic results. For the field quenches we have added the square of the rescaled modulus of the magnetization. We observe that within the numerical error its value and the correlation function for large $d$ coincide as predicted. Comparing the decay of the correlations for the thermal system to the decay in the quenched system we find for the interaction quenches a good agreement both for the $4\times4$ and the $12\times12$ system. The curves show only small deviations and have the same shape going to $0$ for large distances $d$. The rt-VMC results show a very good agreement to the exact results apart from the interaction quench $(0;3.5)\rightarrow(1;3.5)$. For the field quenches the differences between the time-averaged results after the quenches and the thermal results are more significant. For small quenches the shapes of the curves are very similar and the values coincide within the numerical error. For larger quenches we still observe a good agreement between the rt-VMC results and the exact results, but here larger deviations from the thermal curves arise. Although the absolute differences between the curves for a given value of $d$ are not too large, the decay of the correlations as function of $d$ is different for strong field quenches. After the field quenches the correlations decay faster with $d$ than for the system in thermal equilibrium and the stationary long-range value of the correlations is reached at larger distances $d$ and is lower than for the thermal system.

\section*{Related work}
Recently an exact theorem on generalized thermalization in $D$-dimensional quantum systems in the thermodynamic limit has been formulated \cite{Doyon2016}. The theorem states that generalized thermalization can be observed if the state of the system is algebraically sizably clustering. It also holds for exponentially sizably clustering states. Then the stationary state of the system can be described by a GGE which has to take into account all local and quasilocal charges of the system. For non-integrable systems, for which the total energy is the only conserved quantity, the generalized thermalization reduces to thermalization with a stationary state according to the CGE. We now discuss the exact theorem with respect to the 2D-TFIM. Considering the 2D-TFIM one has to distinguish between the ferromagnetic and the paramagnetic phase. In the paramagnetic phase the 2D-TFIM is gapped and there is no symmetry breaking both for finite system sizes as well as in the thermodynamic limit. As the ground state of gapped quantum systems is sizably exponentially clustering \cite{Hastings2004,Nachtergaele2006} the exact theorem can be applied to the 2D-TFIM in the thermodynamic limit after the interaction quenches in the paramagnetic phase and predicts thermalization. In our numerical studies we have indeed observed a very good agreement both between the time-averaged observables and their thermal counterparts as well as between the distributions for small quenches, i.e. large ratios $h/J$, also for the finite system sizes that we can simulate. For larger quenches closer to the phase transition we have found deviations between the time averages and the thermal values, but the finite size scaling shows that they decrease with the system size. Our results for the interaction quenches in the paramagnetic phase are thus in agreement with the exact theorem. In the ferromagnetic phase on the other hand the Hamiltonian of the system is not gapped in the thermodynamic limit. The spin flip symmetry is spontaneously broken and long-range order exists, so that all spins of the system are correlated to each other and the correlations do not cluster. An analytic expression for the shape of the decay of the correlations has not been found yet, thus it cannot be decided whether the exact theorem can be applied in the ferromagnetic phase or not.

\section*{Summary and conclusion}
We have studied the quantum relaxation of the 2D-TFIM after global interaction quenches in the paramagnetic phase and global field quenches in the ferromagnetic phase using a newly developed rt-VMC method which allowed us to explore system sizes and time scales that have not been accessible before. In order to answer the question whether this two-dimensional, non-integrable system thermalizes or not we compared time-averaged results after the quenches to results for the system in thermal equilibrium at a temperature defined by the excess energy after the quench. We found that the presence or absence of thermalization depends crucially on the quench protocol and the initial state. For the interaction quenches there is a good agreement between the results for the quenched system and the thermal system in accordance with a recently formulated exact theorem for systems in the thermodynamic limit. Deviations are only observed for strong quenches ending in the vicinity of the phase transition. Finite size scaling suggests that these deviations should vanish in the thermodynamic limit. In contrast to this we have found significant deviations between the thermal results and the time-averaged results after the field quenches in the ferromagnetic phase. These deviations become especially clear comparing the distributions which show deviations already for small quenches for which the thermal expectation values and the time averages still agree. The shape of the distributions indicates that the system does not completely lose the memory of its initial state during the relaxation process, which is a clear contradiction to thermalization. Finite size scaling shows that the deviations do not decrease with the system size either. Although we currently cannot give an explanation for the observed deviations between the thermal system and the time averages after the field quenches in the ferromagnetic phase, we assume that they might be related to the long-range order or the structure of the spectrum of the Hamiltonian without an energy gap between the ground state and the first excited state.

\section*{Appendix}
\subsection*{Ansatz function in the ferromagnetic phase}
The unitary time evolution of the state of the system after the quench reads in terms of the eigenbasis of $\hat{H}_{\text{f}}$
\begin{align}
\ket{\Psi(t)}=\sum_{\lambda}e^{\imath E_{\text{f},\lambda}t}\braket{\Psi_{\text{f},\lambda}|\Psi_{\text{i},0}}\ket{\Psi_{\text{f},\lambda}}
\label{Eq:time_evolution}
\end{align}
with $\ket{\Psi_{\text{f},\lambda}}$ the eigenstate of $\hat{H}_{\text{f}}$ to the eigenvalue $E_{\text{f},\lambda}$. The $\ket{\Psi_{\text{f},\lambda}}$ and $E_{\text{f},\lambda}$ are a priori unknown. Due to the structure of $\ket{\Psi_{\text{i},0}}$ for field quenches starting from $h_{\text{i}}=0$ according to equation (\ref{Eq:GS_field_quench}), $\braket{\Psi_{\text{f},\lambda}|\Psi_{\text{i},0}}$ is proportional to the sum of the coefficients of the two completely ordered basis states in the representation of $\ket{\Psi_{\text{f},\lambda}}$ in the $\mathbf{x}$-basis. Thus $\braket{\Psi_{\text{f},\lambda}|\Psi_{\text{i},0}}$ decreases with increasing energy $E_{\text{f},\lambda}$ and vanishes for antisymmetric eigenstates of $\hat{H}_{\text{f}}$.\\
We use equation (\ref{Eq:time_evolution}) to derive a variational ansatz for $\ket{\Psi(t)}$ for the rt-VMC in the ferromagnetic phase. For vanishing transverse field $h$ the states of the $\mathbf{x}$-basis are eigenstates of the Hamiltonian. Their energies are determined just by the number $n$ of kinks. Turning on the transverse field, the Hamiltonian is not diagonal in the $\mathbf{x}$-basis any more. The effect of the non-diagonal part of the Hamiltonian is to flip the orientation of spins. For small system sizes we computed the eigenstates of the Hamiltonian with exact diagonalization and found that for small ratios $h/J$ basis states of the $\mathbf{x}$-basis with the same number $n$ of kinks and the same magnetization, which we determine by the number $m$ of spin down, have in good approximation the same coefficients, so that
\begin{align}
\ket{\Psi_{\text{f},\lambda}}\approx\sum_{m,n}c_{m,n}^{\text{f},\lambda}\ket{\Psi_{m,n}}
\label{Eq:approximation_ferromagnetic}
\end{align}
with $\ket{\Psi_{m,n}}$ according to equation (\ref{Eq:ansatz_ferro}). The accuracy of the approximation (\ref{Eq:approximation_ferromagnetic}) decreases with increasing ratio $h/J$. For this reason we only apply it to field quenches starting from the completely ordered state and not leaving the ferromagnetic phase. Another important point for the accuracy of the ansatz is the spatial dimension $D$ of the system, as for a fixed number $N$ of sites the number of symmetries within the system increases in higher dimensions. With increasing number of symmetries the number of basis states which can be transformed into each other by symmetry transformations like translation, rotation or reflection, also increases. For basis states which can be transformed into each other by symmetry transformations the description with the same variational parameter is exact.\\
Using the approximation in equation (\ref{Eq:approximation_ferromagnetic}) we can rewrite equation (\ref{Eq:time_evolution}):
\begin{align}
\ket{\Psi(t)}\approx\sum_{m,n}\underbrace{\sum_{\lambda}e^{\imath E_{\text{f},\lambda}t}\braket{\Psi_{\text{f},\lambda}|\Psi_{\text{i},0}}c_{m,n}^{\text{f},\lambda}}_{\displaystyle\alpha_{m,n}(t)}\ket{\Psi_{m,n}}\;.
\label{Eq:ansatz_ferromagnetic}
\end{align}
The $\alpha_{m,n}(t)$ are the variational parameters of the rt-VMC for the field quenches in the ferromagnetic phase.\\
As shown in the main part of the text by a comparison to results of exact time evolution for the $4\times4$ system, using the ansatz function (\ref{Eq:ansatz_ferromagnetic}) for the rt-VMC calculations allows an accurate description of the time evolution of the 2D-TFIM after a field quench starting from the completely ordered state.

\subsection*{Time evolution after field quenches}
The exact time evolution of the variational state in equation (\ref{Eq:ansatz_ferro}) according to the Schr{\"o}dinger equation reads
\begin{align}
\ket{\dot{\Psi}(t)}_{\text{exact}}=-\imath\hat{H}_{\text{f}}\ket{\Psi(t)}
\label{Eq:Psi_dot_exact}
\end{align}
and its variational time dynamics
\begin{align}
\ket{\dot{\Psi}(t)}_{\text{var}}=\sum_{m,n}\frac{\dot{\alpha}_{m,n}(t)}{\sqrt{N_{m,n}}}\sum_{k}\ket{\Psi_{m,n}^{k}}\;.
\label{Eq:Psi_dot_var}
\end{align}
Thus inserting equations (\ref{Eq:Psi_dot_exact}) and (\ref{Eq:Psi_dot_var}) into equation (\ref{Eq:Euclidian_distance}) we find for the Euclidian distance
\begin{align}
\mathcal{D}(t)=\sum_{m',n',k'}\bigg|-\imath\sum_{m,n}\frac{\alpha_{m,n}(t)}{\sqrt{N_{m,n}}}\sum_{k}\braket{\Psi_{m',n'}^{k'}|\hat{H}_{\text{f}}|\Psi_{m,n}^{k}}-\frac{\dot{\alpha}_{m',n'}(t)}{\sqrt{N_{m',n'}}}\bigg|^{2}\;.
\end{align}
Obviously $\mathcal{D}(t)$ is minimal if each summand is $0$, i.e. if
\begin{align}
\frac{\dot{\alpha}_{m',n'}(t)}{\sqrt{N_{m',n'}}}=-\imath\sum_{m,n}\frac{\alpha_{m,n}(t)}{\sqrt{N_{m,n}}}\sum_{k}\braket{\Psi_{m',n'}^{k'}|\hat{H}_{\text{f}}|\Psi_{m,n}^{k}}\;.
\end{align}
Summation over $k'$, multiplication with $\imath$ and division through $\sqrt{N_{m',n'}}$ leads to the equations of motion of the variational parameters in equation (\ref{Eq:equations_of_motion_ferromagnetic}).

\newpage

\addtocounter{page}{-5}



\section*{Supplementary material}

# - Supplementary material -
# Test of quantum thermalization in the two-dimensional transverse-field Ising model

**Benjamin Blaß**[1,*] **and Heiko Rieger**[1,†]

[1]Theoretical Physics, Saarland University, 66123 Saarbrücken, Germany
*bebla@lusi.uni-sb.de
†h.rieger@mx.uni-saarland.de

## ABSTRACT

In this supplementary material we describe the determination of the coefficients in the equations of motion of the variational parameters after the interaction and the field quenches. In case of the interaction quenches the coefficients are time-dependent expectation values which are evaluated at each time step with a single spin flip quantum Monte Carlo algorithm described in the first section. In contrast to this our ansatz for the field quenches in the ferromagnetic phase leads to a set of coupled linear differential equations of first order with constant (time-independent) coefficients which are independent of the actual quench parameters. Their determination is thus a pure combinatorics problem solved with rare event sampling described in the second section.

### Coefficients of the equations of motion for the interaction quenches

The coefficients in the equations of motion of the variational parameters after the interaction quenches are time-dependent expectation values given by

$$\langle \hat{\mathscr{O}} \rangle_t = \frac{\sum_{\mathbf{x}} |\Psi(\mathbf{x},t)|^2 \mathscr{O}(\mathbf{x})}{\sum_{\mathbf{x}} |\Psi(\mathbf{x},t)|^2} \;. \quad (1)$$

We approximate them by mean values using a single spin flip Monte Carlo algorithm[1]. The statistical error of the mean values is inverse proportional to the square root of the number of samples, so that for a sufficiently large number of samples the mean values are good approximations for the expectation values.

The algorithm starts in a random configuration $\mathbf{x}_0$ of the system, i.e. the orientation of each spin (*up* or *down*) is randomly chosen. In each Monte Carlo step the flip of one single spin of the system is proposed. The acceptance probability $A(\mathbf{x} \to \mathbf{x}')$ for the transition from configuration $\mathbf{x}$ to $\mathbf{x}'$ reads

$$A(\mathbf{x} \to \mathbf{x}',t) = \min\left[1, Q(\mathbf{x} \to \mathbf{x}',t)\right] \quad \text{with} \quad Q(\mathbf{x} \to \mathbf{x}',t) = \frac{P(\mathbf{x}',t)T(\mathbf{x}' \to \mathbf{x})}{P(\mathbf{x},t)T(\mathbf{x} \to \mathbf{x}')} \;. \quad (2)$$

$P(\mathbf{x},t)$ is the probability distribution of the states of the $\mathbf{x}$-basis, which is defined by the variational state at the time $t$ according to

$$P(\mathbf{x},t) = \frac{|\langle \mathbf{x}|\Psi(t) \rangle|^2}{\sum_{\mathbf{x}'} |\langle \mathbf{x}'|\Psi(t) \rangle|^2} = \frac{|\Psi(\mathbf{x},t)|^2}{\sum_{\mathbf{x}'} |\Psi(\mathbf{x}',t)|^2} \;. \quad (3)$$

As the operators $\hat{C}^{xx}_{\mathbf{r}}$ are diagonal in the $\mathbf{x}$-basis, the scalar product $\Psi(\mathbf{x},t)$ can be easily computed:

$$\Psi(\mathbf{x},t) = \frac{1}{\sqrt{2^N}} \exp\left(\sum_{\mathbf{r}} \alpha_{\mathbf{r}}(t) C^{xx}_{\mathbf{r}}(\mathbf{x})\right) \quad \text{with} \quad C^{xx}_{\mathbf{r}}(\mathbf{x}) = \langle \mathbf{x}|\hat{C}^{xx}_{\mathbf{r}}|\mathbf{x} \rangle \;. \quad (4)$$

The summation runs over all independent directions $\mathbf{r}$ in the lattice. Two directions are called independent, if they cannot be transformed into each other by exchanging components or altering their signs. There are up to 8 dependent directions belonging to $\mathbf{r}$. If $\mathbf{r} = (r_x, r_y)$ these directions are $(r_x, r_y)$, $(r_x, -r_y)$, $(-r_x, r_y)$, $(-r_x, -r_y)$, $(r_y, r_x)$, $(r_y, r_x)$, $(r_y, -r_x)$, $(-r_y, r_x)$ and $(-r_y, -r_x)$. Their number reduces if two or more of the aforementioned vectors are identical. The number of all spin pairs

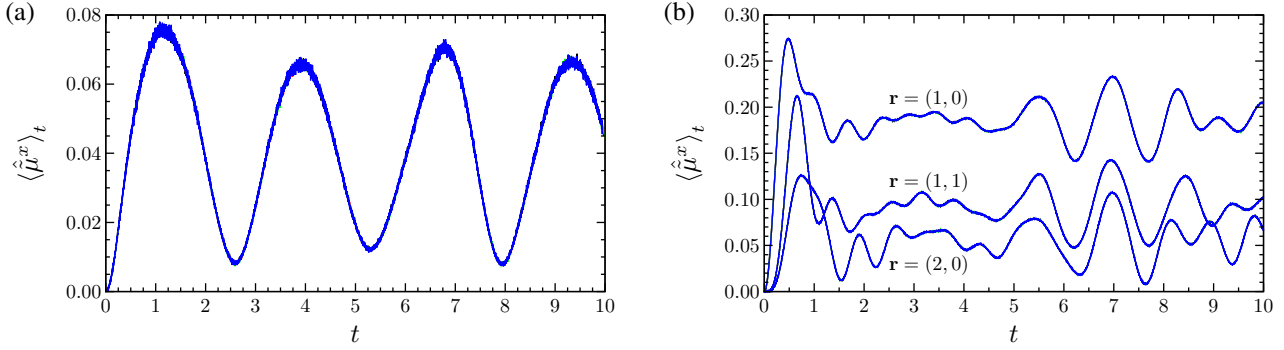

**Figure 1.** Time evolution of (a) the rescaled modulus of the magnetization and (b) the equal time correlation function for different **r** after the interaction quench $(0;3.5) \to (1;3.5)$ in the $16 \times 16$ system. The colour code is as follows: — nmcs = 20000, sbm = $2N$, $\delta t = 0.01$ / — nmcs = 40000, sbm = $2N$, $\delta t = 0.01$ / — nmcs = 20000, sbm = $4N$, $\delta t = 0.01$ / — nmcs = 20000, sbm = $2N$, $\delta t = 0.001$. nmcs defines the number of samples for the Monte Carlo average, sbm the number of Monte Carlo steps between two measurements and $\delta t$ the stepwidth for the numerical integration with the Runge-Kutta scheme. We observe a perfect agreement between the curves.

of the systems whose distance belongs to the independent direction **r** is denoted with $N_\mathbf{r}$.

$T(\mathbf{x} \to \mathbf{x}')$ is the sampling distribution function. For the 2D-TFIM it is time-independent and inverse proportional to the number $K(\mathbf{x})$ of basis states which can be reached from the current configuration via the off-diagonal element of the Hamiltonian. $K(\mathbf{x})$ is thus for each configuration $\mathbf{x}$ equal to the number $N$ of sites in the system.

Inserting the results for $P(\mathbf{x},t)$ and $T(\mathbf{x} \to \mathbf{x}')$ into equation (2), we find

$$A(\mathbf{x} \to \mathbf{x}',t) = \min\left[1, \exp\left\{2\sum_\mathbf{r} \alpha_\mathbf{r}^R(t)\left(C_\mathbf{r}^{xx}(\mathbf{x}') - C_\mathbf{r}^{xx}(\mathbf{x})\right)\right\}\right] \quad (5)$$

with $\alpha_\mathbf{r}^R(t)$ the real part of $\alpha_\mathbf{r}(t)$. The acceptance probability $A(\mathbf{x} \to \mathbf{x}')$ of a proposed Monte Carlo step thus does not depend explicitly on the Hamiltonian of the system, but the parameters of the Hamiltonian just enter via the time-evolved state defining the probability distribution.

The initial configuration $\mathbf{x}_0$ of the system, i.e. the orientation of each individual spin, is randomly chosen. For this configuration the relative orientations of all possible spin pairs are determined and used to compute the correlation functions $C_\mathbf{r}^{xx}(\mathbf{x}_0)$ for each independent direction **r**. In each Monte Carlo step the flip of a randomly chosen spin of the system is proposed. Let its position be **R**. To compute the acceptance probability $A(\mathbf{x} \to \mathbf{x}')$ for the transition from the configuration $\mathbf{x}$ to the proposed configuration $\mathbf{x}'$, which differs from $\mathbf{x}$ only by the inverted orientation of the spin at the position **R**, the correlation functions $C_\mathbf{r}^{xx}(\mathbf{x}')$ of the proposed configuration $\mathbf{x}'$ are needed. They can be efficiently computed flipping the spin at position **R** and then computing the relative orientations between the flipped spin and all other spins of the system. In this way the $C_\mathbf{r}^{xx}(\mathbf{x}')$ can be computed from the already known $C_\mathbf{r}^{xx}(\mathbf{x})$. If the step is accepted, the configuration $\mathbf{x}$ is replaced by $\mathbf{x}'$ and the correlation functions $C_\mathbf{r}^{xx}(\mathbf{x})$ by the $C_\mathbf{r}^{xx}(\mathbf{x}')$.

As the initial configuration is randomly chosen, we wait $10N$ Monte Carlo steps for the system to equilibrate before the first measurement. Measurement means that the expectation values of the operators in the current configuration of the system are used in the computation of the mean values. After the equilibration measurements are done each $2N$ Monte Carlo steps so that the generated samples are independent. We do 20000 measurements, which we split up in 20 independent Monte Carlo runs with different seeds and 1000 samples each. The approximation of the expectation value is thus the average over 20 Monte Carlo runs. The time propagation, i.e. the solution of the equations of motion after the determination of their coefficients, is done with a fourth order Runge-Kutta scheme with a stepwidth of 0.01.

To check the convergence of the Monte Carlo algorithm and its accuracy we used different seeds for the Monte Carlo runs, i.e. different initial configurations and different sequences, and compared the results. In addition we changed the number of Monte Carlo steps between two measurements to check that the samples used to compute the mean values are really independent. We then changed the number of samples for the Monte Carlo averages to reduce their error. Finally we reduced the stepwidth of the Runge-Kutta scheme. We found that for the above values of the parameters of the Monte Carlo algorithm the results are robust with respect to the described tests. This is illustrated in Figure 1 for the rescaled modulus of the magnetization and three different equal time correlation functions of the $16 \times 16$ system (the largest system size we consider) and the quench $(0;3.5) \to (1;3.5)$ (the strongest interaction quench we consider). Using the aforementioned parameters the CPU time for the $16 \times 16$ system is approximately 400 seconds for each time step on one core of an Intel Xeon CPU E5-2680 v2 running at

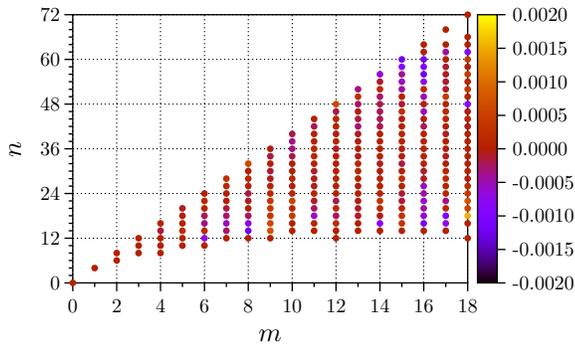

**Figure 2.** Relative error of the results of the RES with respect to the exact results for the $N_{m,n}$ for a $6 \times 6$ system. The values are mean values of 10 Monte Carlo runs of $2 \cdot 10^8$ Monte Carlo samples each. Values for $m > \frac{N}{2}$ are symmetric with respect to $m = \frac{N}{2}$.

2.8 GHz. This includes the single spin flip Monte Carlo algorithm and the numerical integration with the Runge-Kutta scheme. The Monte Carlo algorithm to compute the coefficients in the equations of motion of the variational parameters scales with $\mathcal{O}(N^2)$ as there are $2N$ Monte Carlo steps between two measurements and in each Monte Carlo step the relative orientation of the spin which is proposed to be flipped to all other spins of the system has to be determined.

## Coefficients of the equations of motion for the field quenches

The determination of the $N_{m,n}$ and $T_{m,n;m',n'}$ to compute the $t_{m,n;m',n'}$ in the equations of motion of the variational parameters after the field quenches is a pure combinatorics problem, in which each spin configuration has the same probability given by the inverse dimension of the Hilbert space. As there are large differences between the dimensions $N_{m,n}$ of the subspaces $(m,n)$, standard Monte Carlo methods are not able to reach low-dimensional subspaces in an efficient way and return reliable results. For this reason we apply rare event sampling (RES)[2]. Its idea is to first generate samples according to the original distribution, the majority of which will belong to subspaces of a large dimension. Then the sampling distribution is changed so that its maximum is shifted to subspaces for which only a small number of samples is generated according to the original distribution. As the results of the unbiased and the biased sampling have to agree, the estimators for the subspaces of small dimensions can be determined choosing the a priori unknown partition function under the constraint to minimize the difference between the results of the unbiased sampling and the biased sampling for regions of the Hilbert space in which both results are reliable. This procedure is repeated until the whole Hilbert space has been covered iteratively.

The $N_{m,n}$ and $T_{m,n;m',n'}$ are determined separately for each value of $m$. We compute them as mean values over several independent Monte Carlo sequences with different seeds and different initial states and use the ratio of the standard deviation and the mean value to decide whether the results are reliable or not. Each Monte Carlo sequence consists of one unbiased Monte Carlo run and several biased Monte Carlo runs. The number of biased Monte Carlo runs is a priori unknown, as during the Monte Carlo sequence new Monte Carlo runs with a changed sampling distributions are started until reliable results for all possible values of $n$ with respect to $m$ have been obtained. Each Monte Carlo run consists of a certain number of Monte Carlo steps, each of which proposes the interchange of the position of one spin up and one spin down, thus not altering $m$ but only $n$. The number $n_{\text{new}}$ of kinks in the proposed configuration is computed from the number $n_{\text{old}}$ of kinks in the current configuration. In order to do so only the nearest neighbours of the spin down and the spin up whose positions are proposed to be interchanged have to be considered to determine how the number of kinks in the system would be changed by the proposed Monte Carlo step. For the unbiased sampling each Monte Carlo step is accepted. In the biased sampling a potential $V(n)$ and an artificial inverse temperature $\beta$ are introduced. $V(n)$ depends on a parameter $n_0$, which shifts the position of the maximum of the distribution. $\beta$ changes the width of the distribution. The acceptance probability of a step from a configuration with $n_{\text{old}}$ kinks to a configuration with $n_{\text{new}}$ kinks is for the biased sampling

$$p_{\text{accept}} = \min\left\{1, e^{-\beta \Delta V}\right\} \quad \text{with} \quad \Delta V = V(n_{\text{new}}) - V(n_{\text{old}}). \tag{6}$$

The potential $V(n)$ can be chosen freely, as in the last step of a Monte Carlo run the results according to the unbiased distribution are determined. Thus the focus in the choice of $V(n)$ lies on a high efficiency of the algorithm, which is mainly a question of experience. After tests with different potentials we have chosen $V(n)$ in the following way: Let $n_m^\star$ be the most probable

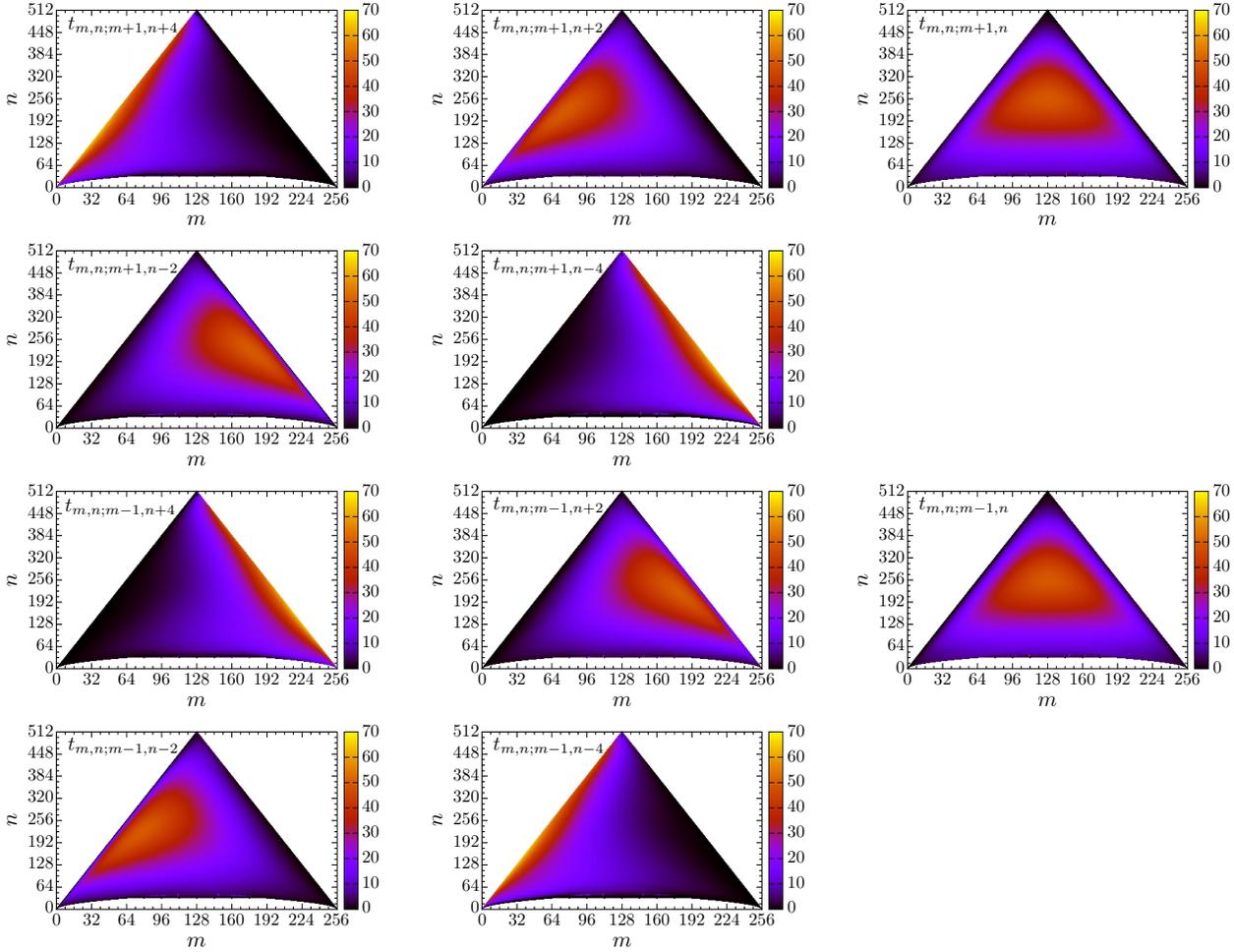

**Figure 3.** Coefficients $t_{m,n;m',n'}$ for the equations of motion of the variational parameters after the field quenches for the $16 \times 16$ system. One observes $t_{m,n;m+1,n+\Delta n} = t_{m,n;m-1,n-\Delta n}$.

number of kinks for a given value of $m$ according to the unbiased distribution. We then use

$$V(n) = \begin{cases} 0 & \text{if } n \leq n_0 \\ (n-n_0)^2 & \text{else} \end{cases} \quad (7)$$

to sample $N_{m,n}$ and $T_{m,n;m',n'}$ for $n < n_m^\star$ and

$$V(n) = (n-n_0)^2 \quad (8)$$

for $n > n_m^\star$.

$\beta$ and $n_0$ are changed after each Monte Carlo run to shift the distribution further into regions of the Hilbert space for which results are still missing in the following way: If in the last Monte Carlo run no new reliable results have been obtained, $\beta$ is increased to narrow the distribution. Otherwise $\beta$ is kept constant and $n_0$ is set just below or above the smallest or largest $n$ respectively for which the results have been accepted. This procedure is repeated until for the considered value of $m$ for all allowed values of $n$ the results for $N_{m,n}$ and $T_{m,n;m',n'}$ have been accepted.

We checked our results for the $N_{m,n}$ and $T_{m,n;m',n'}$ in different ways. First we used the algorithm to compute the coefficients for the system sizes $4 \times 4$ and $6 \times 6$, for which $N_{m,n}$ and $T_{m,n;m',n'}$ can also be determined exactly analyzing each state of the Hilbert space. In Figure 2 we show exemplarily the relative error of the $N_{m,n}$ determined for the $6 \times 6$ system. We observe a very good agreement with deviations only slightly larger than one per mill. The results for the $T_{m,n;m',n'}$ are of the same accuracy. For larger system sizes we checked the validity of our results increasing the number of Monte Carlo runs per sequence until the




\begin{thebibliography}{99}
\bibitem{Cazalilla2006}
Cazalilla, M. A. Effect of Suddenly Turning on Interactions in the Luttinger Model. \textit{Phys. Rev. Lett.} \textbf{97}, 156403 (2006).

\bibitem{Rigol2007}
Rigol, M., Dunjko, V., Yurovsky, V. \& Olshanii, M. Relaxation in a Completely Integrable Many-Body Quantum System: An Ab Initio Study of the Dynamics of the Highly Excited States of 1D Lattice Hard-Core Bosons. \textit{Phys. Rev. Lett.} \textbf{98}, 050405 (2007).

\bibitem{Rigol2008}
Rigol, M., Dunjko, V. \& Olshanii, M. Thermalization and its mechanism for generic isolated quantum systems. \textit{Nature} \textbf{452}, 854 (2008).

\bibitem{Rigol2009}
Rigol M. Breakdown of Thermalization in Finite One-Dimensional Systems. \textit{Phys. Rev. Lett.} \textbf{103}, 100403 (2009).

\bibitem{Manmana2007}
Manmana, S. R., Wessel, S., Noack, R. M. \& Muramatsu, A. Strongly Correlated Fermions after a Quantum Quench. \textit{Phys. Rev. Lett.} \textbf{98}, 210405 (2007).

\bibitem{Kollath2007}
Kollath, C., L{\"a}uchli, A. M. \& Altman, E. Quench Dynamics and Nonequilibrium Phase Diagram of the Bose-Hubbard Model. \textit{Phys. Rev. Lett.} \textbf{98}, 180601 (2007).

\bibitem{Calabrese2007}
Calabrese, P. \& Cardy, J. Quantum quenches in extended systems. \textit{J. Stat. Mech. Theor. Exp.} P06008 (2007).

\bibitem{Fagotti2008}
Fagotti, M. \& Calabrese, P. Evolution of entanglement entropy following a quantum quench: Analytic results for the $XY$ chain in a transverse magnetic field. \textit{Phys. Rev. A} \textbf{78}, 010306 (2008).

\bibitem{Sotiriadis2009}
Sotiriadis, S., Calabrese, P. \& Cardy, J. Quantum quench from a thermal initial state. \textit{Europhys. Lett.} \textbf{87}, 20002 (2009).

\bibitem{Calabrese2011}
Calabrese, P., Essler, F. H. L. \& Fagotti, M. Quantum Quench in the Transverse-Field Ising Chain. \textit{Phys. Rev. Lett.} \textbf{106}, 227203 (2011).

\bibitem{Calabrese2012}
Calabrese, P., Essler, F. H. L. \& Fagotti, M. Quantum quench in the transverse field Ising chain: I. Time evolution of order parameter correlators. \textit{J. Stat. Mech. Theor. Exp.} P07016 (2012).

\bibitem{Calabrese2012a}
Calabrese, P., Essler, F. H. L. \& Fagotti, M. Quantum quenches in the transverse field Ising chain: II. Stationary state properties. \textit{J. Stat. Mech. Theor. Exp.} P07022 (2012).

\bibitem{Moeckel2008}
Moeckel, M. \& Kehrein, S. Interaction Quench in the Hubbard Model. \textit{Phys. Rev. Lett.} \textbf{100}, 175702 (2008).

\bibitem{Cramer2008}
Cramer, M., Dawson, C. M., Eisert, J. \& Osborne, T. J. Exact Relaxation in a Class of Nonequilibrium Quantum Lattice Systems. \textit{Phys. Rev. Lett.} \textbf{100}, 030602 (2008).

\bibitem{Barmettler2009}
Barmettler, P., Punk, M., Gritsev, V., Demler, E. \& Altman, E. Relaxation of Antiferromagnetic Order in Spin-$1/2$ Chains Following a Quantum Quench. \textit{Phys. Rev. Lett.} \textbf{102}, 130603 (2009).

\bibitem{Rossini2009}
Rossini, D., Silva, A., Mussardo, G. \& Santoro, G. Effective Thermal Dynamics Following a Quantum Quench in a Spin Chain. \textit{Phys. Rev. Lett.} \textbf{102}, 127204 (2009).

\bibitem{Schiro2010}
Schir{\'o}, M. \& Fabrizio, M. Time-Dependent Mean Field Theory for Quench Dynamics in Correlated Electron Systems \textit{Phys. Rev. Lett.} \textbf{105}, 076401 (2010).

\bibitem{Igloi2000}
Igl{\'o}i, F. \& Rieger, H. Long-Range Correlations in the Nonequilibrium Quantum Relaxation of a Spin Chain. \textit{Phys. Rev. Lett.} \textbf{85}, 3233 (2000).

\bibitem{Igloi2011}
Igl{\'o}i, F. \& Rieger, H. Quantum Relaxation after a Quench in Systems with Boundaries. \textit{Phys. Rev. Lett.} \textbf{106}, 035701 (2011).

\bibitem{Rieger2011}
Rieger, H. \& Igl{\'o}i, F. Semiclassical theory for quantum quenches in finite transverse Ising chains. \textit{Phys. Rev. B} \textbf{84}, 165117 (2011).

\bibitem{Blass2012}
Bla{\ss}, B., Rieger, H. \& Igl{\'o}i, F. Quantum relaxation and finite-size effects in the XY chain in a transverse field after global quenches. \textit{Europhys. Lett.} \textbf{99}, 30004 (2012).

\bibitem{Caux2013}
Caux, J.-S. \& Essler, F. H. L. Time Evolution of Local Observables After Quenching to an Integrable Model. \textit{Phys. Rev. Lett.} \textbf{110}, 257203 (2013).

\bibitem{Khatami2013}
Khatami, E., Pupillo, G., Srednicki, M. \& Rigol, M. Fluctuation-Dissipation Theorem in an Isolated System of Quantum Dipolar Bosons after a Quench. \textit{Phys. Rev. Lett.} \textbf{111}, 050403 (2013).

\bibitem{Bucciantini2014}
Bucciantini, L., Kormos, M. \& Calabrese, P. Quantum quenches from excited states in the Ising chain. \textit{J. Phys. A: Math. Theor.} \textbf{47} 175002 (2014).

\bibitem{Fagotti2014}
Fagotti, M., Collura, M., Essler, F. H. L. \& Calabrese, P. Relaxation after quantum quenches in the spin-$\frac{1}{2}$ Heisenberg XXZ chain. \textit{Phys. Rev. B} \textbf{89}, 125101 (2014).

\bibitem{Heyl2015}
Heyl, M. Scaling and Universality at Dynamical Quantum Phase Transitions. \textit{Phys. Rev. Lett.} \textbf{115}, 140602 (2015).

\bibitem{James2015}
James, A. J. A. \& Konik, R. M. Quantum quenches in two spatial dimensions using chain array matrix product states. \textit{Phys. Rev. B} \textbf{92}, 161111(R) (2015).

\bibitem{Strand2015}
Strand, H. U. R., Eckstein, M. \& Werner, P. Nonequilibrium Dynamical Mean-Field Theory for Bosonic Lattice Models. \textit{Phys. Rev. X} \textbf{5}, 011038 (2015).

\bibitem{Essler2016}
Essler, F. H. L. \& Fagotti, M. Quench dynamics and relaxation in isolated integrable quantum spin chains. \textit{J. Stat. Mech. Theor. Exp.} P064002 (2016).

\bibitem{Reimann2008}
Reimann, P. Foundation of Statistical Mechanics under Experimentally Realistic Conditions. \textit{Phys. Rev. Lett.} \textbf{101}, 190403 (2008).

\bibitem{Barthel2008}
Barthel, T. \& Schollw{\"o}ck, U. Dephasing and the Steady State in Quantum Many-Particle Systems. \textit{Phys. Rev. Lett.} \textbf{100}, 100601 (2008).

\bibitem{Biroli2010}
Biroli, G., Kollath, C. \& L{\"a}uchli, A. M. Effect of Rare Fluctuations on the Thermalization of Isolated Quantum Systems \textit{Phys. Rev. Lett.} \textbf{105}, 250401 (2010).

\bibitem{Eckstein2008}
Eckstein, M. \& Kollar, M. Nonthermal Steady States after an Interaction Quench in the Falicov-Kimball Model. \textit{Phys. Rev. Lett.} \textbf{100}, 120404 (2008).

\bibitem{Eckstein2009}
Eckstein, M., Kollar, M. \& Werner, P. Thermalization after an Interaction Quench in the Hubbard Model. \textit{Phys. Rev. Lett.} \textbf{103}, 056403 (2009).

\bibitem{Tsuji2013}
Tsuji, N., Eckstein, M. \& Werner, P. Nonthermal Antiferromagnetic Order and Nonequilibrium Criticality in the Hubbard Model. \textit{Phys. Rev. Lett.} \textbf{110}, 136404 (2013).

\bibitem{Marcuzzi2013}
Marcuzzi, M., Marino, J., Gambassi, A. \& Silva, A. Prethermalization in a Nonintegrable Quantum Spin Chain after a Quench. \textit{Phys. Rev. Lett.} \textbf{111}, 197203 (2013).

\bibitem{Sirker2014}
Sirker, J., Konstantinidis, N. P., Andraschko, F. \& Sedlmayr, N. Locality and thermalization in closed quantum systems. \textit{Phys. Rev. A} \textbf{89}, 042104 (2014).

\bibitem{Gogolin2011}
Gogolin, C., M{\"u}ller, M. P. \& Eisert, J. Absence of Thermalization in Nonintegrable Systems. \textit{Phys. Rev. Lett.} \textbf{106}, 040401 (2011).

\bibitem{Riera2012}
Riera, A., Gogolin, C. \& Eisert, J. Thermalization in Nature and on a Quantum Computer. \textit{Phys. Rev. Lett.} \textbf{108}, 080402 (2012).

\bibitem{Eisert2015}
Eisert, J., Friesdorf, M. \& Gogolin, C. Quantum many-body systems out of equilibrium \textit{Nature Physics} \textbf{11}, 124 (2015).

\bibitem{Gogolin2016}
Gogolin, C. \& Eisert, J. Equilibration, thermalisation, and the emergence of statistical mechanics in closed quantum systems. \textit{Rep. Prog. Phys.} \textbf{79}, 056001 (2016).

\bibitem{Cassidy2011}
Cassidy, A. C., Clark, C. W. \& Rigol, M. Generalized Thermalization in an Integrable Lattice System. \textit{Phys. Rev. Lett.} \textbf{106}, 140405 (2011).

\bibitem{Caux2012}
Caux, J.-S. \& Konik, R. M. Constructing the Generalized Gibbs Ensemble after a Quantum Quench. \textit{Phys. Rev. Lett.} \textbf{109}, 175301 (2012).

\bibitem{Mussardo2013}
Mussardo, G. Infinite-Time Average of Local Fields in an Integrable Quantum Field Theory After a Quantum Quench. \textit{Phys. Rev. Lett.} \textbf{111}, 100401 (2013).

\bibitem{Pozsgay2013}
Pozsgay, B. The generalized Gibbs ensemble for Heisenberg spin chains. \textit{J. Stat. Mech. Theor. Exp.} P07003 (2013).

\bibitem{Brockmann2014}
Brockmann, M. et al. Quench action approach for releasing the N{\'e}el state into the spin-1/2 XXZ chain. \textit{J. Stat. Mech. Theor. Exp.} P12009 (2014).

\bibitem{Wouters2014}
Wouters, B. et al. Quenching the Anisotropic Heisenberg Chain: Exact Solution and Generalized Gibbs Ensemble Predictions. \textit{Phys. Rev. Lett.} \textbf{113}, 117202 (2014).

\bibitem{Goldstein2014}
Goldstein, G. \& Andrei, N. Failure of the local generalized Gibbs ensemble for integrable models with bound states. \textit{Phys. Rev. A} \textbf{90}, 043625 (2014).

\bibitem{Pozsgay2014}
Pozsgay, B. et al. Correlations after Quantum Quenches in the $XXZ$ Spin Chain: Failure of the Generalized Gibbs Ensemble. \textit{Phys. Rev. Lett.} \textbf{113}, 117203 (2014).

\bibitem{Pozsgay2014a}
Pozsgay, B. Failure of the generalized eigenstate thermalization hypothesis in integrable models with multiple particle species. \textit{J. Stat. Mech. Theor. Exp.} P09026 (2014).

\bibitem{Pozsgay2014b}
Pozsgay, B. Quantum quenches and generalized Gibbs ensemble in a Bethe Ansatz solvable lattice model of interacting bosons. \textit{J. Stat. Mech. Theor. Exp.} P10045 (2014).

\bibitem{Essler2015}
Essler, F. H. L., Mussardo, G. \& Panfil, M. Generalized Gibbs ensembles for quantum field theories. \textit{Phys. Rev. A} \textbf{91}, 051602 (2015).

\bibitem{Ilievski2015}
Ilievski, E. et al. Complete Generalized Gibbs Ensembles in an Interacting Theory. \textit{Phys. Rev. Lett.} \textbf{115}, 157201 (2015).

\bibitem{Ilievski2016}
Ilievski, E., Medenjak, M., Prosen, T. \& Zadnik, L. Quasilocal charges in integrable lattice systems. \textit{J. Stat. Mech. Theor. Exp.} P064008 (2016).

\bibitem{Doyon2016}
Doyon, B. Thermalization and pseudolocality in extended quantum systems. \textit{arXiv}:1512.03713.

\bibitem{Larson2013}
Larson, J. Integrability versus quantum thermalization \textit{J. Phys. B: At. Mol. Opt. Phys.} \textbf{46} (2013).

\bibitem{Hamazaki2016}
Hamazaki, R., Ikeda, T. N. \& Ueda, M. Generalized Gibbs ensemble in a nonintegrable system with an extensive number of local symmetries. \textit{Phys. Rev. E} \textbf{93}, 032116 (2016).

\bibitem{Deutsch1991}
Deutsch, J. M. Quantum statistical mechanics in a closed system. \textit{Phys. Rev. A} \textbf{43}, 2046 (1991).

\bibitem{Srednicki1994}
Srednicki, M. Chaos and quantum thermalization \textit{Phys. Rev. E} \textbf{50}, 888 (1994).

\bibitem{Rigol2012}
Rigol, M. \& Srednicki, M. Alternatives to Eigenstate Thermalization. \textit{Phys. Rev. Lett.} \textbf{108}, 110601 (2012).

\bibitem{Fratus2015}
Fratus, K. R. \& Srednicki, M. Eigenstate thermalization in systems with spontaneously broken symmetry. \textit{Phys. Rev. E} \textbf{92}, 040103 (2015).

\bibitem{Mondaini2016}
Mondaini, R., Fratus, K. R., Srednicki, M. \& Rigol, M. Eigenstate thermalization in the two-dimensional transverse field Ising model. \textit{Phys. Rev. E} \textbf{93}, 032104 (2016).

\bibitem{Konstantinidis2015}
Konstantinidis, N. P. Thermalization away from integrability and the role of operator off-diagonal elements. \textit{Phys. Rev. E} \textbf{91}, 052111 (2015).

\bibitem{Konstantinidis2016}
Konstantinidis, N. P. Thermalization of a dimerized antiferromagnetic spin chain. \textit{J. Phys.: Condens. Mat.} \textbf{28}, 026001 (2016).

\bibitem{Chiocchetta2015}
Chiocchetta, A., Tavora, M., Gambassi, A. \& Mitra, A. Short-time universal scaling in an isolated quantum system after a quench. \textit{Phys. Rev. B} \textbf{91}, 220302(R) (2015).

\bibitem{Maraga2015}
Maraga, A., Chiocchetta, A., Mitra, A. \& Gambassi, A. Aging and coarsening in isolated quantum systems after a quench: Exact results for the quantum $\text{O}(N)$ model with $N$ $\ensuremath{\rightarrow}$ $\ensuremath{\infty}$. \textit{Phys. Rev. E} \textbf{92}, 042151 (2015).

\bibitem{Chiocchetta2016}
Chiocchetta, A., Tavora, M., Gambassi, A. \& Mitra, A. Short-time universal scaling and light-cone dynamics after a quench in an isolated quantum system in d spatial dimensions. \textit{Phys. Rev. B} \textbf{94}, 134311 (2016).

\bibitem{Onsager1944}
Onsager, L. Crystal Statistics. I. A Two-Dimensional Model with an Order-Disorder Transition. \textit{Phys. Rev.} \textbf{65}, 117 (1944).

\bibitem{Pfeuty1971}
Pfeuty, P. \& Elliott, R. J. The Ising model with a transverse field. II. Ground state properties. \textit{J. Phys. C: Solid St. Phys.} \textbf{4}, 2370 (1971).

\bibitem{deJongh1998}
du Croo de Jongh, M. S. L. \& van Leeuwen, J. M. J. Critical behavior of the two-dimensional Ising model in a transverse field: A density-matrix renormalization calculation. \textit{Phys. Rev. B} \textbf{57}, 8494 (1998).

\bibitem{Rieger1999}
Rieger, H. \& Kawashima, N. Application of a continuous time cluster algorithm to the two-dimensional random quantum Ising ferromagnet. \textit{Eur. Phys. J. B} \textbf{9}, 233 (1999).

\bibitem{Binder1981}
Binder, K. Finite Size Scaling Analysis of Ising Model Block Distribution Functions. \textit{Z. Phys. B} \textbf{43}, 119 (1981).

\bibitem{Pfeuty1970}
Pfeuty, P. The One-Dimensional Ising Model with a Transverse Field. \textit{Ann. Phys.} \textbf{57}, 79 (1970).

\bibitem{Fradkin1989}
Fradkin, E. Jordan-Wigner transformation for quantum-spin systems in two dimensions and fractional statistics. \textit{Phys. Rev. Lett.} \textbf{63}, 322 (1989).

\bibitem{Wang1991}
Wang, Y. R. Ground state of the two-dimensional antiferromagnetic Heisenberg model studied using an extended Wigner-Jordon transformation. \textit{Phys. Rev. B} \textbf{43}, 3786(R) (1991).

\bibitem{Azzouz1993}
Azzouz, M. Interchain-coupling effect on the one-dimensional spin-1/2 antiferromagnetic Heisenberg model. \textit{Phys. Rev. B} \textbf{48}, 6136 (1993).

\bibitem{Carleo2012}
Carleo, G., Becca, F., Schir{\'o}, M. \& Fabrizio, M. Localization and Glassy Dynamics Of Many-Body Quantum Systems. \textit{Sci. Rep.} \textbf{2}, 243 (2012).

\bibitem{Carleo2014}
Carleo, G., Becca, F., Sanchez-Palencia, L., Sorella, S. \& Fabrizio, M. Light-cone effect and supersonic correlations in one- and two-dimensional bosonic superfluids. \textit{Phys. Rev. A} \textbf{89}, 031602(R) (2014).

\bibitem{Cevolani2015}
Cevolani, L., Carleo, G. \& Sanchez-Palencia, L. Protected quasilocality in quantum systems with long-range interactions. \textit{Phys. Rev. A} \textbf{92}, 041603(R) (2015).

\bibitem{Ido2015}
Ido, K., Ohgoe, T. \& Imada, M. Time-dependent many-variable variational Monte Carlo method for nonequilibrium strongly correlated electron systems. \textit{Phys. Rev. B} \textbf{92}, 245106 (2015).

\bibitem{Bishop2000}
Bishop, R. F., Farnell, D. J. J. \& Ristig, M. L. Ab Initio Treatments of the Ising Model in a Transverse Field. \textit{Int. J. Mod. Phys. B} \textbf{14}, 1517 (2000).

\bibitem{Hartmann2002}
Hartmann, A. K. Sampling rare events: Statistics of local sequence alignments. \textit{Phys. Rev. E} \textbf{65}, 056102 (2002).

\bibitem{Hastings2004}
Hastings, M. B. Locality in Quantum and Markov Dynamics on Lattices and Networks. \textit{Phys. Rev. Lett.} \textbf{93}, 140402 (2004).

\bibitem{Nachtergaele2006}
Nachtergaele, B. \& Sims, R. Lieb-Robinson bounds and the exponential clustering theorem. \textit{Commun. Math. Phys.} \textbf{265}, 119 (2006).

\bibitem{Kliesch2014}
Kliesch, M., Gogolin, C., Kastoryano, M. J., Riera, A. \& Eisert, J. Locality of Temperature. \textit{Phys. Rev. X} \textbf{4}, 031019 (2014).
\end{thebibliography}
\end{document}